\documentclass[fleqn]{article}

\usepackage{amsmath,amssymb}
\usepackage{graphicx}
\usepackage[top=0.75in, bottom=0.75in, left=0.75in, right=0.75in, dvips]{geometry}

\begin{document}

\title{\sf Optimal Renormalization-Group Improvement
of Two Radiatively-Broken Gauge Theories}

\title{\sf Optimal Renormalization-Group Improvement
of Two Radiatively-Broken Gauge Theories}

\author{V. Elias,$^{1,2}$ R.B. Mann,$^{1,3}$ D.G.C. McKeon,
$^{2}$ and T.G. Steele$^{4}$}
\footnotetext[1]{
Perimeter Institute for Theoretical Physics, 35 King Street North,
Waterloo, ON  N2J 2W9,  Canada}
\footnotetext[2]{
Department of Applied Mathematics, The University of Western Ontario,
London, ON  N6A 5B7, Canada}
\footnotetext[3]{
Department of Physics, University of Waterloo, Waterloo, ON  N2L 3G1,
Canada}
\footnotetext[4]{
Department of Physics \& Engineering Physics, University of Saskatchewan,
Saskatoon, SK  S7N 5E2, Canada}

\maketitle

\begin{abstract}
In the absence of a tree-level scalar-field mass, renormalization-group
(RG) methods permit the explicit summation of leading-logarithm
contributions to all orders of the perturbative series for the
effective-potential functions utilized in radiative symmetry breaking. For scalar-field
electrodynamics, such a summation of leading logarithm contributions leads to upper bounds on the
magnitudes of both gauge and scalar-field coupling constants, and suggests the possibility of an
additional phase of spontaneous symmetry breaking characterized by a scalar-field mass comparable
to that of the theory's gauge boson. For radiatively-broken electroweak symmetry, the all-orders
summation of leading logarithm terms involving the dominant three couplings (quartic scalar-field,
$t$-quark Yukawa, and QCD) contributing to standard-model radiative corrections leads to an
RG-improved potential characterized by a  $216$ {\rm GeV} Higgs boson mass. Upon incorporation of electroweak
gauge couplants we find that the predicted Higgs mass increases to $218\,{\rm GeV}$.
The potential is also characterized by a quartic scalar-field coupling
over five times larger than that anticipated for an equivalent Higgs
mass obtained via conventional spontaneous symmetry breaking, leading to a concomitant enhancement of processes (such as $W^+ W^- \rightarrow ZZ$) sensitive to this coupling.
Moreover, if the QCD coupling constant is taken to
be sufficiently strong, the tree potential's local minimum at  $\phi = 0$ is shown to be restored for
the summation of leading logarithm corrections. Thus if QCD exhibits a two-phase structure similar
to that of $N = 1$ supersymmetric Yang-Mills theory, the weaker asymptotically-free phase of QCD
{\it may be selected} by the large logarithm behaviour of the RG-improved effective potential for radiatively
broken electroweak symmetry.
\end{abstract}

\section{Introduction:  Radiatively Broken Abelian Gauge Symmetry}

In their 1973 paper \cite{A}, S. Coleman and E. Weinberg demonstrated that spontaneous symmetry
breaking occurs within gauge theories in which scalar fields are initially massless.
This approach to symmetry breaking has considerable predictive power and led to perhaps
the first definitive prediction for the magnitude of the Higgs boson mass. Although this
prediction has since proved to be incorrect (the mass of the top-quark  was unknown at
the time of their paper), Coleman and Weinberg's work also demonstrated the nontrivial
role radiative corrections play in determining observable consequences of gauge theories,
with eventual applications to cosmology and empirical standard model physics.

The simplest example of calculable radiatively induced symmetry breaking
considered by Coleman and Weinberg is that of massless scalar electrodynamics, in
which an initially massless complex scalar field (or alternatively, its two constituent
real-field components) is minimally coupled to an unbroken Abelian gauge theory.

The effective potential of this massless scalar electrodynamics is generated from the
tree-level potential

\renewcommand{\theequation}{1.\arabic{equation}}
\setcounter{equation}{0}

\begin{equation}
V = \lambda (\phi_1^2 + \phi_2^2)^2 / 24
\label{eq1.1}
\end{equation}
by the scalar field self-interaction and the interaction Lagrangian
involving real scalar $\phi_1$, $\phi_2$, and gauge fields $A_\mu$:
\begin{equation}
{\cal{L}} = \frac{1}{2} \left( \partial_\mu \phi_1 - e A_\mu \phi_2
\right)^2 + \frac{1}{2} \left( \partial_\mu \phi_2 + e A_\mu \phi_1
\right)^2  - V.
\label{eq1.2}
\end{equation}
The effective potential for this theory is calculated in Landau gauge in Ref.\ \cite{A},
\begin{equation}
V_{eff}= \phi^4 \left[ \frac{\lambda}{24} + \left(
\frac{5\lambda^2}{1152\pi^2} + \frac{3e^4}{64\pi^2} \right) \left( \log
\frac{\phi^2}{\mu^2} + k \right) + {\cal{O}} \left( \lambda^3, e^6
\right) \right],
\label{eq1.3}
\end{equation}
and in arbitrary covariant gauge in Appendix A. The absence of an
explicit scalar field mass term precludes the need for a cosmological
constant term in $V_{eff}$. The renormalization constant $k$ is determined
by the (choice-of-scheme) definition of
the quartic scalar interaction constant $\lambda$ as the fourth
derivative of the effective potential with respect to the classical
field $\phi_{1c}$ (or $\phi_{2c}, \; \phi_{1c}^2 + \phi_{2c}^2 \equiv
\phi_c^2$) when evaluated at the renormalization mass $\mu$:
\begin{equation}
\left. \frac{d^4 V}{d \phi_{1c}^4}\right|_{\phi_c^2 = \mu^2} \equiv \lambda
\label{eq1.4}
\end{equation}
in which case
\begin{equation}
k = - 25/6.
\label{eq1.5}
\end{equation}
The condition that $dV_{eff} / d \phi_c = 0$ at the vacuum expectation
value $\langle \phi \rangle$, in conjunction with the assumption that $\lambda$ and $e^4$ are of
equivalent order, leads to the constraint
\begin{equation}
\lambda = 33 e^4 / 8\pi^2.
\label{eq1.6}
\end{equation}
Given this constraint, one finds that
\begin{equation}
V_{eff} = \frac{3e^4}{64\pi^2} \phi_c^4 \left[ \log
\frac{\phi_c^2}{\langle\phi\rangle^2} - \frac{1}{2} \right] + {\cal{O}}(e^6) .
\label{eq1.7}
\end{equation}
The scalar field and gauge field mass terms [the RG-invariance 
of the former is discussed in Appendix B] are respectively given by
\begin{equation}
m_\phi^2 = V_{eff}''(\langle\phi\rangle), \; \; \; m_A^2 = e^2
\langle\phi\rangle^2 ,
\label{eq1.8}
\end{equation}
and one finds from Eq.\ (\ref{eq1.7}) that \cite{A}	
\begin{equation}
\frac{m_{\phi}^2}{m_A^2} = 3e^2/8\pi^2.
\label{eq1.9}
\end{equation}

The result (\ref{eq1.9}) is obtained entirely via the leading logarithmic
contribution (\ref{eq1.3}) to the scalar field self-coupling and by assuming 
that $\lambda^2$ can be neglected compared to $e^4$. At this juncture, 
if this latter assumption were not
true, then Eq.\ (\ref{eq1.6}) would no longer be true, and $\lambda$ could be
sufficiently large to render order $\lambda^3$-and-higher contributions
to Eq.\ (\ref{eq1.3}) too important to neglect.  In Section V of Ref.\ \cite{A},
renormalization group methods were utilized to show that the range of
validity of the above results [particularly Eq.\ (\ref{eq1.9})] could be extended to
arbitrary but still small values of $\lambda$ and $e$.  The form of
renormalization group improvement employed in Ref.\ \cite{A} is the introduction of
running coupling constants in the effective potential; it is then argued
that $\lambda$ can be moved from any (small) value to an
${\cal{O}}(e^4)$ value via a change in renormalization mass $\mu$ for which $e$
retains a small value.

In the following sections we employ a more ``optimal'' form of
renormalization-group (RG) improvement in which the leading logarithms
of the effective potential for massless scalar electrodynamics are
explicitly summed to {\it all} orders in perturbation theory.  Indeed, we show in
the next section that the logarithmic contribution to the effective
potential (\ref{eq1.3}) is just the first term of such a summation of leading
logarithms, and that one can obtain this term directly from the 
renormalization-group equation (RGE), rather than via the explicit calculation 
and summation of one loop diagrams (as in Ref.\ \cite{A}).  Our methodological point of 
view (as articulated in general terms
by Maxwell \cite{B}) is to incorporate {\it all} information about higher order
contributions to the effective potential that is accessible via RG
methods.  Such all-orders summations of logarithms, which are compared to
more conventional forms of RG-improvement in refs. \cite{C,D}, have been previously applied to
the effective potential of $\phi^4$ scalar field theory \cite{E,F,Y,Z}, various effective
actions \cite{G}, as well as to correlation functions, decay rates and
cross-sections for a number of perturbatively-calculated processes
\cite{C}.

In the present paper, we re-examine both the massless scalar electrodynamics (${\rm M}\!\!\!$/SED)
and the radiatively-broken $SU(2)\times U(1)$ electroweak symmetry considered in Ref.\ 1 after 
inclusion of all RG-accessible information about higher order terms within the 
effective potentials of these theories.  In Section 2, we demonstrate how the 
RG-equation for the effective potential of ${\rm M}\!\!\!$/SED leads to recursion relations 
which serve to determine {\it all} leading logarithm contributions to the perturbative 
effective potential series, despite the presence of two perturbative 
couplants, $\lambda/4\pi^2$ and $\alpha/\pi (=e^2 / 4\pi^2)$, in the theory. We restructure 
the theory into a power series in the latter (presumably small) gauge 
couplant by obtaining closed-form expressions in Section 3 for the 
summation-of-leading-logarithm contributions to all orders of the 
self-interaction couplant $\lambda/4\pi^2$ involving a given power of $\alpha/\pi$. 
This procedure reduces a double-summation over powers of both 
couplants to a summation over powers of a single (presumably known) 
gauge couplant $\alpha/\pi$, a summation in which the series coefficients are 
obtained by solving successive first-order differential equations in 
the unknown scalar-field self-interaction couplant $\lambda/4\pi^2$.

In Section 4, we incorporate into the effective potential this entire 
summation-of-leading-logarithm series, as opposed to just the first two terms 
of this series (leading and one-loop) present in Eq.\ (\ref{eq1.3}).  By applying the 
same set of finite renormalization conditions delineated above to the full 
leading-logarithm potential, we are able to obtain results that are not 
limited by assumptions of either couplant being perturbatively small.  
Surprisingly, we find a nonlinear relation between the two couplants of 
the theory that places upper bounds on the magnitudes of both couplants. 
Moreover, the radiative symmetry breaking epitomized by Eq.\ (\ref{eq1.9}) is seen 
to correspond to only the weaker of two possible phases for the scalar-field 
self-interaction couplant. In the stronger phase, the scalar field mass is 
substantially larger, comparable in magnitude to the mass that symmetry-breaking 
generates for the Abelian gauge field.

This latter result provides motivation for a similar re-appraisal of 
the Standard Model for electroweak physics with a single (initially-) 
massless complex scalar field doublet, since the Higgs boson mass is already 
known to be larger than the broken-symmetry $W$ and $Z$ gauge-boson masses for 
$SU(2) \times U(1)$ gauge theory.  In Section 5 we argue that the dominant three 
interaction couplant parameters for this theory are the scalar-field 
self-interaction couplant $\lambda/4\pi^2$, the $t$-quark Yukawa-interaction 
couplant $g_t^2/4\pi^2$, and the QCD gauge-interaction couplant $\alpha_s/\pi$, even 
though this latter couplant does not contribute to leading-logarithms 
within the scalar-field effective potential until two-loop order. With 
this simplification, the RGE for the effective potential leads 
to recursion relations from which leading-logarithm contributions may be 
extracted to arbitrary order in all three couplants, including, of 
course, the one-loop Yukawa and scalar-field couplant contributions 
obtained in Ref.\ \cite{A}. 

In Section 6 we reorganize perturbative leading-logarithm contributions 
to the effective potential as a power series in the Yukawa couplant 
$g_t^2 / 4\pi^2$, whose series coefficients are closed-form all-orders functions 
of the remaining two couplants. These functions are determined from a set 
of successive partial differential equations. The closed form solutions for the first 
three of these series coefficients are obtained through explicit solution 
of such partial differential equations. [The solution of the fourth series 
coefficient is presented in Appendix C.]

It is shown in Section 7 that any prediction of the Higgs boson mass is 
sensitive to terms with at most four powers of the logarithm appearing 
in the effective potential's perturbative series. In anticipation of 
this result, Section 6 lists every such term arising from contributions 
of the three dominant couplants to the leading-logarithms of that series. 
In Section 7, these results are utilized to obtain a prediction of 216 {\rm GeV} 
for the Higgs boson mass, a result that follows from a nonlinear relationship 
between the scalar-field self-interaction couplant and the known QCD 
gauge- and $t$-quark Yukawa couplants. Moreover, the predicted scalar-field 
couplant is seen to be several times larger than the scalar couplant 
corresponding to an equivalent Higgs boson mass obtained from conventional
spontaneous symmetry-breaking. Thus, given the discovery of a Higgs boson with mass at 
or near 216 {\rm GeV}, a clear signal for radiative symmetry breaking (as opposed 
to conventional spontaneous symmetry breaking) would be the amplification of 
processes such as the $W^+W^- \rightarrow ZZ$ scattering cross-section which exhibit 
significant sensitivity to the scalar-field self-interaction couplant.  

These results are discussed further in Section 8. Residual renormalization-scale 
dependence of the one-loop effective potential is shown to be substantially 
larger than that obtained when leading logarithm contributions are summed. 
Moreover, this minimal residual scale dependence, if indicative of unknown 
subsequent-to-leading-logarithm corrections, buttresses the case for the 
216 {\rm GeV} prediction for the Higgs boson mass to be meaningful.

The summation of leading logarithms, however, is of greatest value in 
ascertaining large-logarithm properties of the effective potential, properties 
corresponding to that potential's large-field and zero-field limits. In Section 8, 
these two limits are examined separately. Large-field contributions to each power
of the Yukawa couplant in the summation-of-leading-logarithms series are shown to
grow singular when $\phi \simeq 22 \langle \phi \rangle$. Since term-by-term singularities
in a series are not necessarily singularities in the function represented by the series,
this ultraviolet singularity represents a bound on the domain of the series obtained
in Section 6, as opposed to a fundamental property of the effective potential itself.
Moreover,  it is shown in Section 8 that every term in the summation-of-leading-logarithms
series diverges {\it positively} as $\phi$ approaches this singularity near
$22 \langle \phi \rangle$ from below. Such a result is consistent with this series representation
(or truncations thereof) being bounded from below over its entire domain of applicability.
The boundedness of the potential prior to an ${\cal O}(5\,{\rm TeV})$ singularity is shown to be confirmed by analysis of a closed-form
exact solution for the summation of leading logarithms that is obtainable in the limit where QCD is turned off ({\it i.e.} the QCD
couplant is set to zero).
   The correspondence between the explicit leading-logarithm summation and
its equivalent method-of-characteristics solution is also demonstrated
in Section 8 for the case of radiatively broken electroweak symmetry
breaking, which facilitates estimation of the contributions of sub-dominant electroweak gauge couplants to
radiative  symmetry breaking.  Such contributions are shown to raise the extracted value of the Higgs boson mass from
$216\,{\rm GeV}$ to $218\,{\rm GeV}$.

We conclude Section 8 by examining this summation-of-leading-logarithms series in the
zero-field limit. Convergence of this series is demonstrated when the QCD couplant is
sufficiently strong ($\alpha_s \geq 0.4$). The series summation for this case is shown to exhibit
a local {\it minimum} at $\phi = 0$, suggestive of electroweak symmetry restoration when the QCD
gauge couplant is sufficiently large. The applicability of this result in distinguishing
between possible coexisting strong and asymptotically-free phases of QCD is briefly discussed.


\section{RG-Equation for the Effective Potential of ${\rm M}\!\!\!$/SED}

\renewcommand{\theequation}{2.\arabic{equation}}
\setcounter{equation}{0}

The effective potential of massless scalar electrodynamics may be
expressed in terms of a perturbative series $S = \frac{\lambda}{4\pi^2}
+ {\cal{O}} \left( \lambda^2, e^4, e^2\lambda \right)$ such that
\begin{equation}
V_{eff} = \frac{\pi^2 \phi^4}{6} S(\lambda, e^2, L)
\label{eq2.1}
\end{equation}
where
\begin{gather}
\phi^2 \equiv \phi_1^2 + \phi_2^2
\label{eq2.2}
\\
L \equiv \log (\phi^2 / \mu^2) .
\label{eq2.3}
\end{gather}
The statement that $V_{eff}$ is independent of the renormalization mass
scale $\mu$ (i.e., that $\mu \; d V_{eff} / d\mu = 0$) implies that
$\lambda$, $e^2$ and $\phi^2$ are all implicit functions of $\mu$ such
that
\begin{equation}
\left\{ \left[ -2 + 2\gamma \right] \frac{\partial}{\partial L} + \beta_e
\frac{\partial}{\partial e^2} + \beta_\lambda \frac{\partial}{\partial
\lambda} + 4\gamma \right\} S(\lambda, e^2, L) = 0
\label{eq2.4}
\end{equation}
where the chain rule coefficients in Eq.\ (\ref{eq2.4}) are just the RG functions \cite{A}
\begin{gather}
\gamma \equiv \frac{\mu}{\phi_1} \frac{d \phi_1}{d\mu} \left( \equiv \frac{\mu}{\phi_2} \frac{d \phi_2}{d\mu}\right)
= \frac{\mu}{2\phi^2} \frac{d\phi^2}{d\mu} = \frac{3 e^2}{16\pi^2} + {\cal{O}} (e^4, \lambda e^2, \lambda^2)
\label{eq2.5}
\\
\beta_e \equiv \mu \frac{d e^2}{d\mu} = \frac{e^4}{24\pi^2} + {\cal{O}} (e^6, \lambda e^4, \lambda^2 e^2, \lambda^3 )
\label{eq2.6}
\\
\beta_\lambda \equiv \mu \frac{d\lambda}{d\mu} = \frac{5}{24\pi^2} \lambda^2 - \frac{3}{4\pi^2} \lambda e^2 + \frac{9}{4\pi^2} e^4
+ {\cal{O}} (\lambda^3, e^2 \lambda^2, e^4 \lambda, e^6).
\label{eq2.7}
\end{gather}
In Eqs.\ (\ref{eq2.5}), (\ref{eq2.6}) and (\ref{eq2.7}), we have listed only the one-loop contributions to $\lambda$, $\beta_e$ and $\beta_\lambda$.  These contributions
are sufficient in themselves to determine leading-logarithm contributions to the series $S(\lambda, e^2, L)$ to all orders
of perturbation theory.  Such leading logarithm contributions to a given order necessarily involve a power of the logarithm $L$ that
is always one less than the aggregate power of the coupling constants $e^2$ and $\lambda$.  The all-orders series of leading logarithm contributions
may be represented in terms of the couplant parameters
\begin{gather}
x(\mu) \equiv e^2 / 4\pi^2
\label{eq2.8}
\\
y(\mu) \equiv \lambda / 4\pi^2
\label{eq2.9}
\end{gather}
as follows:
\begin{equation}
S_{LL} = \sum_{n=1}^\infty \left(R_{n, n-1} \; \; y^n L^{n-1} +
\sum_{k=0}^\infty T_{n, k} \; \; x^n y^k L^{n+k-1} \right).
\label{eq2.10}
\end{equation} 
The only {\it ab initio} known coefficients of this series are $R_{1,0}
= 1$ and $T_{1,0} = 0$, as required to obtain correspondence between Eq.\
(\ref{eq2.10}) and the tree-order $\lambda \phi^4 / 24$ contribution to
$V_{eff}$ (\ref{eq2.1}):
\begin{equation}
S_{LL} = \frac{\lambda}{4\pi^2} + {\cal{O}} \left( \lambda^2, e^4,
\lambda e^2 \right) = y + {\cal{O}} \left( y^2, x^2, xy \right);
\label{eq2.11}
\end{equation}
{\it i.e.}, $T_{1,0}$ must vanish, as there is no $e^2 \phi^4$ tree level
contribution to the potential.  The other coefficients in Eq.\ (\ref{eq2.10}) may be
extracted by considering only those contributions to the RGE (\ref{eq2.4})
which either lower the power of the logarithm $L$ by one, or which raise
the aggregate power of the couplants $x$ or $y$ by one.  Such terms are
entirely known from the one-loop RG functions (\ref{eq2.5}), (\ref{eq2.6}) and (\ref{eq2.7}),
and are seen to lead via Eqs.\ (\ref{eq2.8}) and (\ref{eq2.9}) to the following 
RGE for determining leading-logarithm coefficients:
\begin{equation}
\left[-2\frac{\partial}{\partial L} +\left( \frac{5}{6} y^2 - 3xy +
9x^2 \right) \frac{\partial}{\partial y} + \frac{x^2}{6}
\frac{\partial}{\partial x} + 3x \right] S_{LL} (x, y, L) = 0.
\label{eq2.12}
\end{equation}

Note that this equation is sufficient in itself to determine the
one-loop effective potential.  If we substitute the series (\ref{eq2.10}) into
Eq.\ (\ref{eq2.12}), we find that the aggregate coefficient of $y^2$ vanishes
provided
\begin{equation}
-2 R_{2,1} + \frac{5}{6} R_{1,0} = 0.
\label{eq2.13}
\end{equation}
Since $R_{1,0} = 1$, as argued above, we see that $R_{2,1} = 5/12$.  The
aggregate coefficient of $xy$ in Eq.\ (\ref{eq2.12}) vanishes provided $T_{1,1} =
0$, which explains the absence of a $\lambda e^2 \log \phi^2$ ``cross
term'' in the one-loop effective potential.  Similarly, the aggregate
coefficient of $x^2$ in Eq.\ (\ref{eq2.12}) vanishes provided 
\begin{equation}
-2 T_{2,0} + 9 R_{1,0} + \frac{19}{6} T_{1,0} = 0 .
\label{eq2.14}
\end{equation}
Since $T_{1,0} = 0$ and $R_{1,0} = 1$, we see that $T_{2,0} = 9/2$.
Using the series (\ref{eq2.10}) to obtain only the (one-loop) contributions 
linear in the (leading) logarithm $L$, we find that
\begin{equation}
\begin{split}
V_{LL} & =  \frac{\pi^2}{6} \phi^4 S_{LL}
 =  \frac{\pi^2}{6} \phi^4
\left( R_{1,0} y + T_{1,0} x + R_{2,1} y^2 L + T_{1,1} xyL + T_{2,0} x^2
L + {\cal{O}} (L^2) \right)
\\
& =  \frac{\pi^2}{6} \phi^4 \left( y + \frac{5}{12} y^2 L + \frac{9}{2}
x^2 L + {\cal{O}}(L^2)\right)
\\
& =  \frac{\lambda \phi^4}{24} + \left( \frac{5\lambda^2}{1152\pi^2} +
\frac{3 e^2}{64\pi^2} \right) \left( \phi^4 \log \left( \frac{\phi^2}{\mu^2}
\right) \right) + {\cal{O}} \left[ \log^2 \frac{\phi^2}{\mu^2} \right].
\end{split}
\label{eq2.15}
\end{equation}
The result (\ref{eq2.15}) is, of course, the same as Coleman and Weinberg's
\cite{A} direct one-loop calculation quoted in Eq.\ (\ref{eq1.3}).  The
remaining term in Eq.\ (\ref{eq1.3}) is just the finite $- \frac{25}{6} \left(
\frac{5\lambda^2}{1152\pi^2} + \frac{3 e^4}{64\pi^2} \right) \phi^4$
counterterm required by the $d^4 V / d\phi^4|_\mu = \lambda$
renormalization condition, as discussed in the previous section.  Of
course, any RG approach, such as that leading to Eq.\ (\ref{eq2.15}), ultimately
relies on Feynman diagrammatic calculations of the RG functions (\ref{eq2.5}),
(\ref{eq2.6}) and (\ref{eq2.7}).  It is nevertheless reassuring that these one-loop RG
functions lead, via RG methods, to the same one-loop effective potential
as one obtains explicitly from Feynman diagrams with external scalar field
legs.\footnote{An early analysis of radiative spontaneous symmetry breaking
using the RG equation also appears in Ref.\ \cite{ZZ}.}

However, the result (\ref{eq2.15}) does not include all information available
from the leading-logarithm RGE (\ref{eq2.12}).  If we substitute Eq.\ (\ref{eq2.10})
into Eq.\ (\ref{eq2.12}), we obtain recursion relations which determine
{\it all} coefficients $R_{n, n-1}$ and $T_{n,k}$ in the series (\ref{eq2.10})
for leading-logarithm contributions to $V_{eff}$.  For arbitrary power
$p \geq 2$, we find that the aggregate coefficient of $y^p L^{p-2}$
vanishes in Eq.\ (\ref{eq2.12}) provided
\begin{equation}
-2 (p-1) R_{p, p-1} + \frac{5}{6} (p-1) R_{p-1, p-2} = 0 .
\label{eq2.16}
\end{equation}
Since $R_{1,0} = 1$, we see from this constraint that $R_{p, p-1} = \left(
\frac{5}{12} \right)^{p-1}$.  Similarly, we find that the aggregate
coefficients of $xy^p L^{p-1}$ and $x^2 y^p L^p$ vanish provided
\begin{gather}
-2p T_{1,p} + \frac{5}{6} (p-1) T_{1, p-1} - 3p R_{p, p-1} + 3 R_{p, p-
1} = 0, \; \; \; p \geq 1
\label{eq2.17}
\\
 -  2 (p+1) T_{2, p} + \frac{5}{6} (p-1) T_{2, p-1} - 3p T_{1,p} 
 +  9(p+1)R_{p+1, p} + \frac{1}{6} T_{1, p} + 3 T_{1,p} = 0, \; \; \; p \geq 1.
\label{eq2.18}
\end{gather}
Since all coefficients $R_{p, p-1}$ are known from Eq.\ (\ref{eq2.16}), Eq.\ (\ref{eq2.17})
is sufficient to determine all coefficients $T_{1,k}$;  note that the
result $T_{1,1} = 0$ follows directly from Eq.\ (\ref{eq2.17}) with $p = 1$.
Similarly Eqs.\ (\ref{eq2.18}) and (\ref{eq2.14}) are sufficient to determine all
coefficients $T_{2,k}$ in the series (\ref{eq2.10}).  Subsequent coefficients of
terms degree-3-and-higher in $x$ are obtained by demanding that the
aggregate coefficient of $x^n y^p L^{n+p-2}$ vanish:
\begin{gather}
 -  2(p+n-1) T_{n,p} + \frac{5}{6} (p-1) T_{n, p-1} - 3p T_{n-1,
p}
 +  9 (p+1) T_{n-2, p+1} + \frac{n+17}{6} T_{n-1,p} = 0, \; \; \; n \geq 3, \; p \geq 1
\label{eq2.19}
\\
-2(n-1) T_{n,0} + 9 T_{n-2, 1} + \frac{n+17}{6} T_{n-1, 0} =
0, \; \; \; n > 2.
\label{eq2.20}
\end{gather}

Thus, one could in principle use the above set of recursion relations to
determine the {\it entire} leading-logarithm series (\ref{eq2.10}), as opposed
to its one-loop projection (\ref{eq2.15}).  However, we find it most useful to
restructure the series (\ref{eq2.10}) into a series that is perturbative in the
couplant $x$ $(= e^2/4\pi^2)$ but which includes summation over all
powers of $y$ $(= \lambda / 4\pi^2)$,
\begin{equation}
S_{LL} = y S_0 (yL) + x S_1 (y L) + x^2 L S_2 (y L) + \ldots = y S_0 (y L) + \sum_{j=1}^\infty x^j L^{j-1} S_j (y L),
\label{eq2.21}
\end{equation}
since $x$, the electromagnetic couplant $\alpha/\pi$, is anticipated to
be perturbatively small.  If we equate Eqs.\ (\ref{eq2.21}) and (\ref{eq2.10}), we find
that
\begin{gather}
S_0 (y L) = \sum_{n=1}^\infty R_{n, n-1} (y L)^{n-1}
\label{eq2.22}
\\
S_j (y L) = \sum_{k=0}^\infty T_{j, k} (y L)^k.
\label{eq2.23}
\end{gather}
In the next section we will utilize the recursion relations
(\ref{eq2.16}) -- (\ref{eq2.20}) to obtain closed-form expressions for the summations
$S_0$, $S_1$, $S_2$ and $S_3$. Although $S_k$ with $k > 3$ can also be
determined from these recursion relations, we will be able to show (Section 4) 
that the couplant $x$ is constrained by minimization of $V_{eff}$ to be small, in 
which case contributions $x^k L^{k-1} S_k(yL)$ to Eq.\ (\ref{eq2.21}) with $k \geq 3$ can 
be safely disregarded.  Note also from the organization of the series (\ref{eq2.21}) that no
{\it a priori} assumptions are required concerning the magnitude of the
couplant $y$, since all-orders $y$-dependence resides in the closed-form
summations obtained for $S_j$ in the next section.

\section{Summations of Leading Logarithms in ${\rm M}\!\!\!$/SED}

\renewcommand{\theequation}{3.\arabic{equation}}
\setcounter{equation}{0}

The series (\ref{eq2.22}) for $S_0 (y L)$ is just a geometric series, since
$R_{p+1, p} / R_{p, p-1} = 5/12$ by the recursion relation (\ref{eq2.16}).
Since $R_{1,0} = 1$, we find easily that
\begin{equation}
S_0 (y L) = \frac{1}{1 - \frac{5}{12} y L} \equiv \frac{1}{w}.
\label{eq3.1}
\end{equation}
We will find it convenient to make use of the variable $w = 1 -
\frac{5}{12} yL$ to parametrise subsequent summations, and will henceforth
denote by $S_k [w]$ their functional dependence on this variable ({\it i.e.},
$S_0 [w] = 1/w$).  

To find the series $S_1 (u)$, where $u = yL$, we multiply each term of
the recursion relation (\ref{eq2.17}) by $u^{p-1}$ and then sum from $p = 1$ to
$\infty$:
\begin{equation}
- 2 \sum_{p=1}^\infty p T_{1,p} u^{p-1} + \frac{5}{6}
\sum_{p=1}^\infty (p-1) T_{1,p-1} u^{p-1}
- 3 \sum_{p=1}^\infty (p-1) R_{p, p-1} u^{p-1} = 0.
\label{eq3.2}
\end{equation}
We note from the expressions (\ref{eq2.22}) and (\ref{eq2.23}) that
\begin{gather}
S_0 (u) = \sum_{p=1}^\infty R_{p, p-1} u^{p-1},
\label{eq3.3}
\\
S_1 (u) = \sum_{p=1}^\infty T_{1,p-1} u^{p-1}.
\label{eq3.4}
\end{gather}
Consequently Eq.\ (\ref{eq3.2}) is just the first order differential equation
\begin{equation}
-2 \left( 1 - \frac{5}{12} u \right) \frac{dS_1}{du} - 3u
\frac{dS_0}{du} = 0.
\label{eq3.5}
\end{equation}
We change variables to $w = 1 - \frac{5}{12} u$ and, noting from Eq.\
(\ref{eq3.1}) that $S_0 = 1/w$, we find that
\begin{equation}
\frac{dS_1}{dw} = \frac{18}{5} (w^{-3} - w^{-2}),
\label{eq3.6}
\end{equation}
with an initial condition obtained from Eq.\ (\ref{eq3.4}) in the $u \rightarrow
0$ limit:
\begin{equation}
\lim_{w\rightarrow 1} S_1[w] = \lim_{u \rightarrow 0} S_1(u) = T_{1,0}
= 0.
\label{eq3.7}
\end{equation}
The solution to this differential equation is
\begin{equation}
S_1 [w] = -\frac{9}{5} \frac{(w-1)^2}{w^2}
\label{eq3.8}
\end{equation}
where $w = 1 - \frac{5}{12} yL$.

A differential equation for the series
\begin{equation}
S_2 (u) = \sum_{k=0}^\infty T_{2,k} u^k
\label{eq3.9}
\end{equation}
can be obtained by multiplying the recursion relation (\ref{eq2.18}) by $u^p$
and then summing from $p = 1$ to infinity:
\begin{equation}
- 2u \frac{dS_2}{du} - 2 (S_2 - T_{2,0}) + \frac{5}{6} u^2 \frac{d
S_2}{du} - 3u \frac{d S_1}{du}
+ 9u \frac{dS_0}{du} + 9 (S_0 - 1) + \frac{19}{6} S_1 = 0 .
\label{eq3.10}
\end{equation}
The constant terms in Eq.\ (\ref{eq3.10}) cancel;  $T_{2,0} = 9/2$, as obtained from
Eq.\ (\ref{eq2.14}).  If we make the change of variable $w = 1 - \frac{5}{12} u$, we
find that
\begin{equation}
\frac{d S_2}{dw} + \frac{1}{w(w-1)} S_2  =  - \frac{3}{2w}
\frac{dS_1}{dw} + \frac{19}{12w(w-1)} S_1
+ \frac{9}{2w} \frac{d S_0}{dw} + \frac{9}{2w(w-1)} S_0
\label{eq3.11}
\end{equation}
with initial condition
\begin{equation}
\lim_{w \rightarrow 1} S_2[w] = \lim_{u \rightarrow 0} S_2(u) = T_{2,0} =
\frac{9}{2}.
\label{eq3.12}
\end{equation}
Substituting the solutions (\ref{eq3.1}) and (\ref{eq3.8}) for $S_0$ and $S_1$ into the
right hand side of Eq.\ (\ref{eq3.11}), one finds that
\begin{equation}
S_2 [w] = -\frac{1}{20w^3} \left[ -20w^3 - 77w^2 + 34w - 27 \right].
\label{eq3.13}
\end{equation}
A similar analysis of the recursion relations (\ref{eq2.19}) and (\ref{eq2.20}) leads to
the differential equations
\begin{equation}
\frac{dS_k}{dw} + \frac{k-1}{w(w-1)} S_k
= -\frac{3}{2w} \frac{d
S_{k-1}}{dw} + \frac{17 + k}{12w(w-1)} S_{k-1}
- \frac{15}{8w(w-1)} \frac{d S_{k-2}}{dw} \equiv f_k[w]
\label{eq3.14}
\end{equation}
where $w = 1 - 5u/12$ and where $S_k (u)$ is defined by Eq.\ (\ref{eq2.23}).  The
solution to Eq.\ (\ref{eq3.14}) is uniquely determined by the requirement that
$S_k$ not be singular at $w = 1$, since $\stackrel {\lim}{_{w \rightarrow 1}} S_k =
\stackrel{\lim}{_{u \rightarrow 0}} S_k = T_{k,0}$.  Consequently we find from
Eq.\ (\ref{eq3.14}) that
\begin{equation}
S_k = \frac{w^{k-1}}{(w-1)^{k-1}} \int_1^w dr \frac{(r-1)^{k-1}}{r^{k-1}}
f_k[r],
\label{eq3.15}
\end{equation}
where the function $f_k$ is defined to be the inhomogeneous driving term
in Eq.\ (\ref{eq3.14}) obtained from knowledge of $S_{k-1} [w]$ and $S_{k-2} [w]$.
One finds, for example, that
\begin{equation}
S_3 [w] = \frac{1}{240w^4} \left[ 580w^4 + 760w^3 - 323w^2 + 126w - 243 \right]
\label{eq3.16}
\end{equation}
and that $\stackrel{\lim}{_{w \rightarrow 1}} S_3 = T_{3,0}=\frac{15}{4}$,
consistent with Eq.\ (\ref{eq2.20}) [note that $T_{1,1} = 0$ and $T_{2,0} =
9/2$].


\section{Analysis of the Leading-Logarithm ${\rm M}\!\!\!$/SED Effective Potential}

\renewcommand{\theequation}{4.\arabic{equation}}
\setcounter{equation}{0}

The RGE (\ref{eq2.12}) was shown in Section 2 to determine all
coefficients of the leading-logarithm series $S_{LL}$ for the effective
potential (\ref{eq2.1}) of ${\rm M}\!\!\!$/SED.  Using the results
of Section 3, this effective potential may be expressed as follows:
\begin{equation}
V_{eff}^{LL} = \frac{\pi^2 \phi^4}{6} \left[ y S_0 [w] +
\sum_{n=1}^\infty x^n L^{n-1} S_n [w] \right] + K \phi^4
\label{eq4.1}
\end{equation}
where the logarithm $L$ and the couplants $x$ and $y$ ($w = 1 - 5yL/12$)
are respectively defined in Eqs.\ (\ref{eq2.3}), (\ref{eq2.8}) and (\ref{eq2.9}), and where the
series $S_k [w]$ are given explicitly by Eqs, (\ref{eq3.1}), (\ref{eq3.8}), (\ref{eq3.13}) and,
for $k \geq 3$, by Eq.\ (\ref{eq3.15}).  Note that Eq.\ (\ref{eq4.1}) includes a
finite $K \phi^4$ counterterm.  Such a counterterm is also present in
the ``unimproved'' one-loop effective potential (\ref{eq1.3}), as discussed in
Section 1, but the value of this counterterm will be shifted as a result
of the leading logarithm contributions to Eq.\ (\ref{eq4.1}) past one-loop order.

\begin{figure}[htb]
\centering
\includegraphics[scale=0.6]{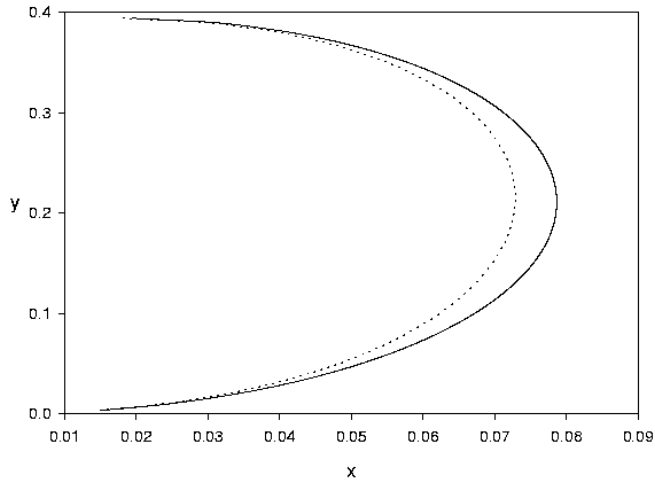}
\caption{
The scalar self-interaction couplant $y$ as a 
double-valued function of the gauge couplant $x$ of ${\rm
M}\!\!\!$/SED.  The solid curve is the solution to the quadratic
equation (\protect\ref{eq4.5}).  The dotted curve is obtained by incorporating all terms
within $V_{eff}^{LL}$ that can contribute to this double-valued relation, 
as discussed in the text.
}
\label{fig1}
\end{figure}

Let us first consider the leading three contributions to Eq.\
(\ref{eq4.1}):
\begin{equation}
V_{eff}^{LL} = \frac{\pi^2 \phi^4}{6} \left[ y S_0 [w] + x S_1 [w] + x^2
L S_2 [w] \right] + K \phi^4 + {\cal{O}} (x^3).
\label{eq4.2}
\end{equation}
We show below that $x$ is constrained by this potential to be small,
providing an {\it a posteriori} justification for truncation of the
series past terms quadratic in $x$.  [The quadratic term in Eq.\ (\ref{eq4.2})
contains the $(3e^4/64\pi^2)\phi^4 L$ term occurring within the one-loop expression
(\ref{eq1.3}).]  As before, the finite $\phi^4$ counterterm is determined by
application of the renormalization condition (\ref{eq1.4}) onto the effective
potential (\ref{eq4.2}).  We then find that
\begin{equation}
\begin{split}
\frac{6 K}{\pi^2}  = & - \left( \frac{125}{72} y^2 + \frac{75}{4} x^2 \right)
- \frac{(2625 y^3 + 1875 y^4 + 625 y^5)}{1296}
\\
& +  \frac{x(175 y^2 + 250 y^3 + 125 y^4)}{48}
- x^2 \frac{(9450 y + 10275 y^2 + 5275 y^3)}{432}
\end{split}
\label{eq4.3}
\end{equation}
Note that the degree two terms in Eq.\ (\ref{eq4.3}) lead to precisely the same finite
counterterm as in Eq.\ (\ref{eq1.3}):
\begin{equation}
K\phi^4 = -\frac{\pi^2}{6} \left( \frac{125 y^2}{72} + \frac{75}{4} x^2 \right) \phi^4 + \ldots
= -\frac{25}{6} \left( \frac{5\lambda^2}{1152 \pi^2} + \frac{3 e^4}{64\pi^2} \right) \phi^4 + \ldots
\label{eq4.4}
\end{equation}
If we substitute Eq.\ (\ref{eq4.3}) into Eq.\ (\ref{eq4.2}), the minimization condition at $\mu^2 = \langle \phi \rangle^2$, 
({\it i.e.}, at $L = 0$) becomes
\begin{equation}
\begin{split}
0  = & \left. \frac{d V_{eff}^{LL}}{d\phi}\right|_{\mu}
= \pi^2 \left[ \left( \frac{1296 y - 1980 y^2 - 2625 y^3 - 1875 y^4 -
    625 y^5}{1944}\right)\right.
\\
& +  x \left( \frac{175 y^2 + 250 y^3 + 125 y^4}{72} \right)
- \left. x^2 \left( \frac{7128 + 9450y + 10275 y^2 + 5275 y^3}{648} \right) \right].
\end{split}
\label{eq4.5}
\end{equation}
Eq.\ (\ref{eq4.5}) is a quadratic equation in $x$ whose solution yields $x$ as a function of $y$ (when $\mu = \langle \phi \rangle$).
The (positive) solutions to this equation are plotted in Fig.\ \ref{fig1}.  We note the following features:

\begin{itemize}
\item[1)]For each value of $x$, there are {\it two} allowed values of $y$.  In other words, for a given value of $e^2$ there are
both a ``strong-$\lambda$'' and ``weak-$\lambda$'' phase of the spontaneously broken theory.
\item[2)]The solution space of Eq.\ (\ref{eq4.5}) places upper bounds on both $x$ and $y$, with a fairly small numerical bound on $x$.
Thus the spontaneously broken theory is truly perturbative in $x$ ({\it i.e.}, in $e^2$), consistent with truncation of 
Eq.\ (\ref{eq4.1}) past its degree-2 terms in $x$.
\item[3)]When $y \stackrel{<}{_\sim} 0.1$, well into the weak-$\lambda$ phase, the solution curve in Fig.\ \ref{fig1}
reduces to $y = 33 x^2 / 2$, consistent with the constraint (\ref{eq1.6}) for the ``unimproved'' one-loop effective potential.
\end{itemize}

\begin{figure}[htb]
\centering
\includegraphics[scale=0.6]{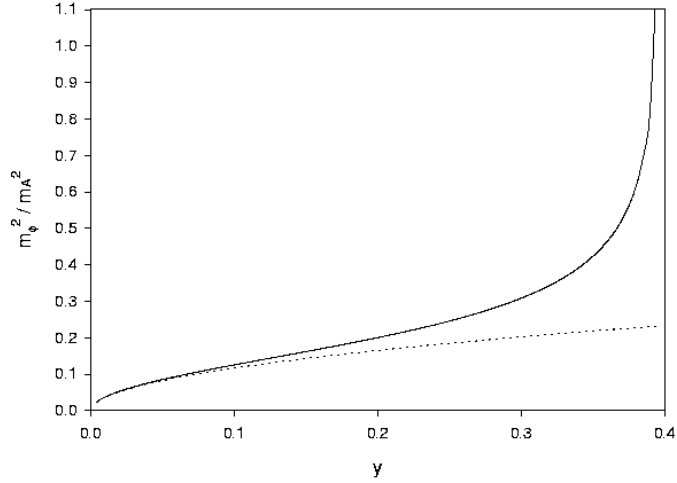}
\caption{
The ratio of the squares of the scalar-field $(\phi)$
and gauge-field $(A)$ masses for ${\rm
M}\!\!\!$/SED as a function of the scalar couplant $y$.  The solid
curve is obtained from Eq.\ (\protect\ref{eq4.6}), with $x(y)$ given by the solution to
Eq.\ (\protect\ref{eq4.5}).  The dotted curve displays corresponding results from the
Coleman-Weinberg relations (\protect\ref{eq1.9}) and (\protect\ref{eq1.6}).
}
\label{fig2}
\end{figure}

\begin{figure}[htb]
\centering
\includegraphics[scale=0.6]{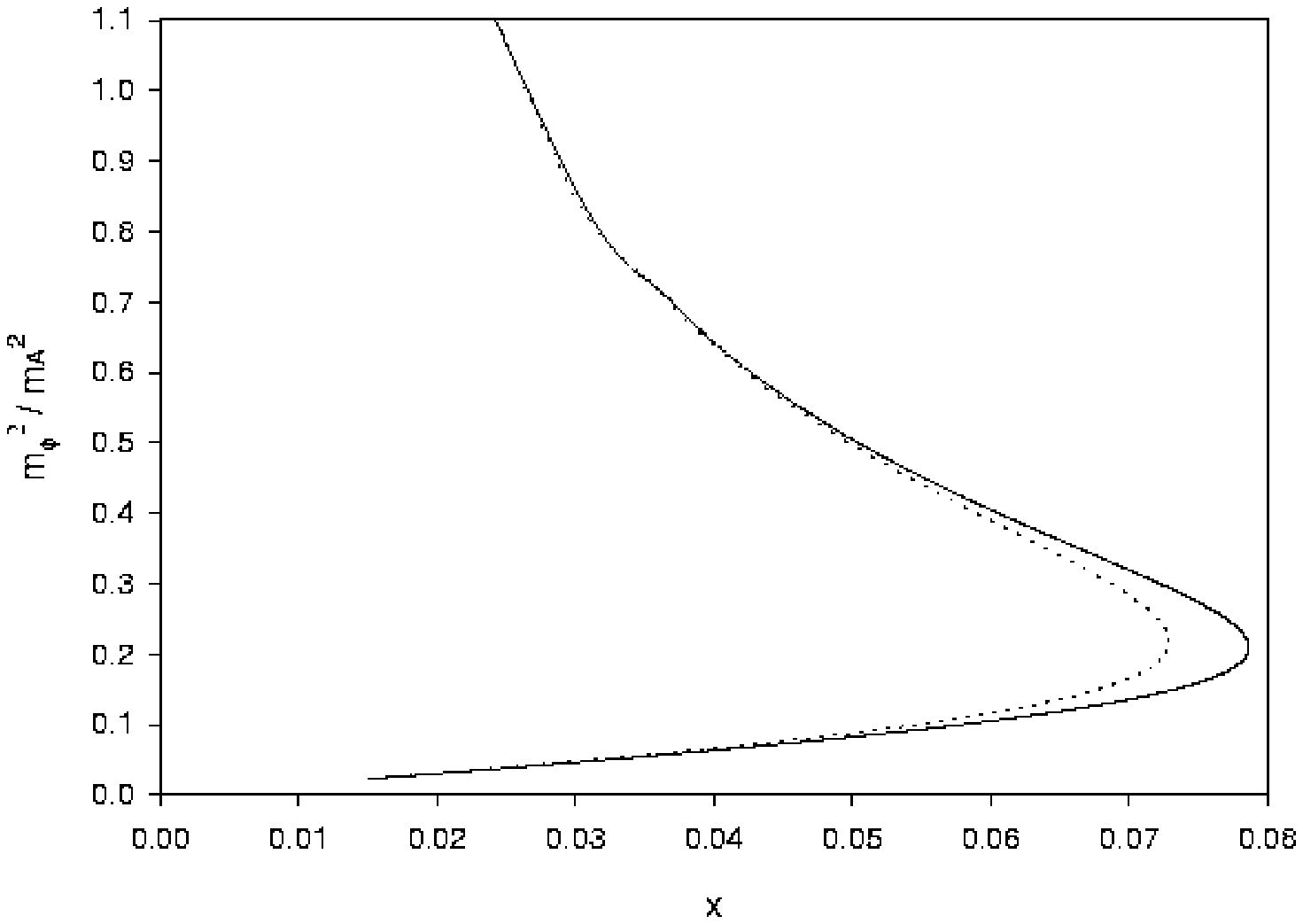}
\caption{
 The dependence of the ratio of squares of the scalar
field and gauge-field masses for ${\rm
M}\!\!\!$/SED on the gauge couplant $x$.  The solid curve is obtained
from Eq.\ (\protect\ref{eq4.6}) with $y(x)$ given implicitly by Eq.\ (\protect\ref{eq4.5}).  The dotted
curve, as in Figure \protect\ref{fig1}, incorporates all additional contributing terms to 
$V_{eff}^{LL}$.
}
\label{fig3}
\end{figure}

The presence of a strong-$\lambda$ phase in the radiatively broken theory suggests the possibility of a large scalar-field
mass solution to the spontaneously broken theory.  The only available scale within the spontaneously broken 
theory for assessing the scalar-field mass $m_\phi$ is the gauge boson mass $m_A = e \langle \phi \rangle$.
Using the effective potential (\ref{eq4.2}) [with counterterm coefficient $K$ given by Eq.\ (\ref{eq4.3})], 
we find using Eqs.\ (\ref{eq1.8}) that
\begin{equation}
\begin{split}
\frac{m_\phi^2}{m_A^2}  = & \left( \frac{y}{2x} - \frac{27x}{4} \right) - \frac{1}{2592x}\left[ 1620 y^2 + 2475 y^3 + 1875 y^4
+ 625 y^5 \right. 
\\ 
& -  \left. x(4455 y^2 + 6750 y^3 + 3375 y^4)
+ x^2 (26730 y + 30825 y^2 + 15825 y^3)\right],
\end{split}
\label{eq4.6}
\end{equation}
where $x$ is the positive solution to the quadratic equation (\ref{eq4.5}).  In the weak-$\lambda$ phase where
$y \cong 33x^2/2$, the leading contribution to Eq.\ (\ref{eq4.6}) is just
\begin{equation}
\frac{m_\phi^2}{m_A^2} = \left( \frac{3x}{2}\right) + \left[ {\cal{O}}(x^3)\right],
\label{eq4.7}
\end{equation}
consistent with Eq.\ (\ref{eq1.9}) for the unimproved one-loop
 effective potential.  In Fig.\ \ref{fig2}
 we plot the solution (\ref{eq4.6}) as a function of $y$ ($=\lambda/4\pi^2$).
The plot shows progressive deviation of the ratio from its one-loop effective-potential value (\ref{eq4.7}) as $y$ increases 
in magnitude.\footnote{Note that $y = 33x^2/2$ in the one-loop potential, in which case $m_\phi^2 / m_A^2 = \sqrt{3y/22}$ by Eq.\ (\ref{eq4.7}).}
As $y$ approaches its upper bound near 0.4, the ratio grows infinite, corresponding to the decoupled radiatively-broken
massless scalar-field theory one would obtain in the limit $e \rightarrow 0, \; m_A = e \langle\phi \rangle \;  \rightarrow 0$.  
The two-phase nature of the spontaneous symmetry breaking is illustrated in Fig.\ \ref{fig3},
where $x$ (instead of $y$) is used as the independent variable in plotting the mass ratio (\ref{eq4.6}).  The figure
shows that for a given value of $x$ in the allowed domain of the constraint (\ref{eq4.5}), there are two values of
the mass ratio $m_\phi^2 / m_A^2$, respectively corresponding to the two allowed values of $y$ (or of $\lambda$) evident in Fig.\ \ref{fig1}.
The weak-$\lambda$ phase is seen to yield a mass ratio quite close to
Coleman and Weinberg's original one-loop prediction (\ref{eq1.9}) for almost the entire allowed domain in $x$.
However, the strong-$\lambda$ phase yields a scalar-boson mass $m_\phi \stackrel {>}{_\sim} 0.45 m_A$ 
that is comparable to or even larger 
[subject to subsequent-to-leading-log corrections] than the gauge boson mass.

Note that the results described above are not contingent upon (or an
artifact of) the truncation of the series within Eq.\ (\ref{eq4.1}) to
terms of degree-2 or less in $x$;  we have performed such a truncation
to provide insight [via Eq.\ (\ref{eq4.5})] into how the double valued
structure of $x(y)$ occurs.  In Section 7 it is shown that only those
series terms degree-4 and less in $L$ contribute to the information we
extract from the effective potential  in 
Figs.\ \ref{fig1}--\ref{fig3}, a consequence following from renormalization conditions [such as Eq.\ (\ref{eq1.4})] that involve $L = 0$ values of at most four derivatives of the effective potential.  The only such terms omitted from Eq.\ (\ref{eq4.2}) are
\begin{gather}
x^3 L^2 S_3 = \frac{15}{4} x^3 L^2 - \frac{5}{6} x^3 y L^3 - \frac{9415}{4 \cdot 12^3} x^3 y^2 L^4 + {\cal{O}}(L^5),
\label{eq4.8}
\\
x^4 L^3 S_4 = 5x^4 L^3 + \frac{2015}{3 \cdot 4^4} x^4 y L^4 + {\cal{O}}(L^5),
\label{eq4.9}
\\
x^5 L^4 S_5 = \frac{65}{48} x^5 L^4 + {\cal{O}}(L^5),
\label{eq4.10}
\end{gather}
results which can be obtained via successive solutions of Eq.\ (\ref{eq3.14}).  The effect of including these new terms within $V_{eff}^{LL}$ is displayed in the additional dashed curves displayed in Figs.\ \ref{fig1} and \ref{fig3}.  Specifically, we find from Fig.\ \ref{fig1} that the upper bound on the couplant $x$ decreases from 0.078 to 0.073, and that the mass ratio curve of Fig.\ \ref{fig3} is ``pulled in'' accordingly.  However, the two-phase structure described above remains evident in these dashed curves, which are now inclusive of {\it all} contributing leading-logarithm effects.


\section{The RGE for Radiative Electroweak Symmetry Breaking}

\renewcommand{\theequation}{5.\arabic{equation}}
\setcounter{equation}{0}

In the absence of an explicit scalar-field mass term, the one-loop (1L) effective potential for 
$SU(2) \times U(1)$ gauge theory is given by \cite{A,I}
\begin{equation}
V_{eff}^{(1L)} = \frac{\lambda \phi^4}{4} + \phi^4 \left[ \frac{12\lambda^2 - 3g_t^2}{64 \pi^2} + 
\frac{3(3g_2^4 + 2g_2^2 g^{\prime \; 2} + g^{\prime \; 4})}{1024 \pi^2} \right]
\left( \log \frac{\phi^2}{\mu^2} - \frac{25}{6} \right)
\label{eq5.1}
\end{equation}
where the $-25/6$ constant is chosen to ensure that
\begin{equation}
\left. \frac{d^4 V_{eff}^{(1L)}}{d\phi^4}\right|_{\mu} = \frac{d^2
  V_{tree}}{d\phi^4} = 6\lambda.
\label{eq5.2}
\end{equation}
There are four distinct coupling constants appearing in Eq.\ (\ref{eq5.1}), the $SU(2)$ coupling constant $g_2$, the $U(1)$ coupling constant 
$g^\prime$, the $t$-quark Yukawa coupling constant $g_t$, and the quartic scalar-field self-interaction coupling constant $\lambda$.
Three of these are known in terms of the electromagnetic coupling $e$,
the weak angle $\theta_w$, and the masses of the $t$-quark and $W$-boson:
\begin{gather}
g_2^2 \equiv e^2 / \sin^2 \theta_w \cong 0.436
\label{eq5.3a}
\\
g^{\prime \; 2} \equiv e^2 / \cos^2 \theta_w \cong 0.127
\label{eq5.3b}
\\
g_t = m_t \sqrt{2} / \langle \phi \rangle = e \; m_t / (\sqrt{2} m_W \sin \theta_w)
\cong 1.00
\label{eq5.3c}
\end{gather}

Prior to the discovery of the $t$-quark, contributions of known-quark
Yukawa couplings $g_q \leq m_q / 175$ {\rm GeV} to Eq.\ (\ref{eq5.1}) could be safely
ignored relative to the contributions of gauge coupling constants,
permitting an analysis very similar to that for massless scalar
electrodynamics.  By assuming $\lambda$ to be order $g_2^4$,
one then obtains a scalar-boson mass $(m_\phi)$ whose magnitude is
perturbatively suppressed \cite{A}:
\begin{equation}
\frac{m_\phi^2}{m_W^2} = \frac{4}{g_2^2 \langle \phi \rangle^2} \left. \frac{d^2
V_{eff}^{(1L)}}{d\phi^2}\right|_{\mu = \langle \phi \rangle}
= \frac{3\alpha(2+ \sec^4 \theta_w)}{8\pi \sin^2 \theta_w} ,
\label{eq5.4}
\end{equation}
a ratio corresponding to a scalar field mass of order $10$ {\rm GeV}.  As in
scalar electrodynamics, such an approach is self-consistent;
optimization of the potential (\ref{eq5.1}) leads to an ${\cal{O}}(g_2^4)$ value
for $\lambda$, so corrections past one-loop order are seen not to
alter appreciably Eq.\ (\ref{eq5.4}). 

Of course, with the discovery of the $t$-quark, the neglect of Yukawa
couplings is no longer justifiable.  Indeed, the $t$-quark's Yukawa
coupling-constant contributions to Eq.\ (\ref{eq5.1}) are large compared to those
of $SU(2) \times U(1)$ gauge coupling constants.  If $g_t = 1$, one
finds that the $V_{eff}^{(1L)\prime}(\langle\phi\rangle)$ $ = 0$ minimization
condition implies that $\lambda \cong 3.6$ and $m_\phi \cong 350$
{\rm GeV}.  This result is questionable, however, because the value
for $\lambda$ may be too large to justify the neglect of
higher-loop contributions.

This failure to obtain a clear prediction for a conformally-invariant
electro-weak-symmetry potential suggests the utility of extending the
one-loop effective potential (\ref{eq5.1}) to include summation over all leading
logarithms, as already considered in massless scalar electrodynamics.
To obtain such a sum, we first note that the RGE for the potential may be
expressed as follows:
\begin{equation}
\begin{split}
0  = & \mu \frac{d}{d\mu} V \left[ \lambda(\mu), g_t (\mu), g_3 (\mu), \phi^2
(\mu), \mu \right] 
\\
 = & \left( \mu \frac{\partial}{\partial\mu} + \beta_\lambda
\frac{\partial}{\partial \lambda} + \beta_t \frac{\partial}{\partial g_t}
+ \beta_3 \frac{\partial}{\partial g_3}
- 2\gamma \phi^2 \frac{\partial}{\partial \phi^2} \right) V \left( \lambda,
g_t, g_3, \phi^2, \mu \right), 
\end{split}
\label{eq5.5}
\end{equation}
where, to one-loop order in $\lambda, g_t$, and the QCD coupling-constant
$g_3$ \cite{I,J}, 
\begin{gather}
\beta_\lambda \equiv \mu \frac{d\lambda}{d\mu} = \frac{48\lambda
g_t^2}{64\pi^2} + \frac{12\lambda^2}{8\pi^2} - \frac{3 g_t^4}{8\pi^2} +
{\cal{O}} \left( \lambda^k g_{3,t}^{6-2k}\right)
\label{eq5.6a}
\\
\beta_t \equiv \mu \frac{d g_t}{d\mu} = \frac{\frac{9}{2} g_t^3 - 8g_t g_3^2}{16\pi^2} +
{\cal{O}}\left( \lambda^k g_{3,t}^{5-2k}\right)
\label{eq5.6b}
\\
\beta_3 \equiv \mu \frac{d g_3}{d\mu} = \frac{-7 g_3^3}{16\pi^2} + {\cal{O}} \left( g_{3,t}^5 \right)
\label{eq5.6c}
\\
\gamma \equiv \frac{-\mu}{\phi} \frac{d\phi}{d\mu} = \frac{3 g_t^2}{16\pi^2} +
{\cal{O}} \left( \lambda^k g_{3,t}^{4-2k} \right).
\label{eq5.6d}
\end{gather}
Since the contribution of the Yukawa coupling-constant $g_t$ to Eq.\ (\ref{eq5.1})
dominates the contributions of the much smaller $SU(2) \times U(1)$
gauge coupling-constants [as evident in Eqs.\ (\ref{eq5.3a})--(\ref{eq5.3c})], we work in the
approximation in which $g$ and $g^\prime$ are equal to zero.  However,
the contribution of the QCD coupling constant to the Yukawa
$\beta$-function (\ref{eq5.6b}) cannot be neglected, since $g_3^2$ is the largest known
coupling constant contributing to the effective potential $\left[ g_3^2
\cong 4\pi \alpha_s (M_z) \cong 1.50 \right]$.

To assess the full leading-logarithm (LL) contribution to the electroweak
effective potential in the absence of an explicit scalar-field mass term,
we utilize the couplant parameters $x$, $y$, $z$
defined at $\mu = \langle \phi \rangle = 2^{-1/4} G_F^{-1/2} \equiv v$:
\begin{gather}
x \equiv g_t^2 (v) / 4\pi^2 \; \; \; (\cong 0.0253)
\label{eq5.7a}
\\
y \equiv \lambda / 4\pi^2
\label{eq5.7b}
\\
z \equiv g_3^2 (v) / 4\pi^2 \; \; \; (\cong 0.0329)
\label{eq5.7c}
\end{gather}
with corresponding one-loop RG-functions derived from Eqs.\ (\ref{eq5.6a})--(\ref{eq5.6d}):
\begin{gather}
\mu \frac{dx}{d\mu} = \frac{9}{4} x^2 - 4xz
\label{eq5.8a}
\\
\mu \frac{dy}{d\mu} = 6y^2 + 3yx - \frac{3}{2} x^2
\label{eq5.8b}\\
\mu \frac{dz}{d\mu} = -\frac{7}{2} z^2
\label{eq5.8c}
\\
\gamma = \frac{3x}{4}
\label{eq5.8d}
\end{gather}
[The value (\ref{eq5.7c}) is obtained from $\alpha_s (M_z) \cong 0.12$ \cite{K}
via evolution of $\alpha_s$ from $M_z$ to $v$.]
As in Section 2, we write the summation-of-leading-logarithms 
effective potential for radiative broken electroweak symmetry (RBEWS) in the form
\begin{equation}
V_{LL} = \pi^2 \phi^4 S_{LL}
= \pi^2 \phi^4 \left\{\sum_{n=0}^\infty x^n
\sum_{k=0}^\infty y^k \sum_{\ell=0}^\infty z^{\ell} C_{n, k, \ell}
L^{n+k+\ell-1} \right\},\; \; (C_{0,0,0} = 0)
\label{eq5.9}
\end{equation}
where the series $S_{LL}$ is the sum of all contributions involving a
power of the logarithm $L\equiv \log (\phi^2 / \mu^2)$ that is only one
degree lower than the aggregate power of the couplants $\{x, y, z \}$.
We keep only those terms in the RGE (\ref{eq5.5}) that either lower the power of
$L$ or raise the aggregate power of couplants by one,
\begin{equation}
\left[ -2 \frac{\partial}{\partial L} + \left( \frac{9}{4} x^2 - 4xz
\right) \frac{\partial}{\partial x} + \left( 6y^2 + 3yx - \frac{3}{2}
x^2 \right) \frac{\partial}{\partial y}
-\frac{7}{2} z^2 \frac{\partial}{\partial z} - 3x \right]
S_{LL} (x,y,z,L) = 0,
\label{eq5.10}
\end{equation}
since such terms (which arise entirely from one-loop RG functions) are
sufficient in themselves to determine all coefficients $C_{n, \ell, k}$
within $S_{LL}$.  Note that the leading contributions to
$S_{LL}$ [Eq.\ (\ref{eq5.9})] are
\begin{equation}
\begin{split}
S_{LL}  = & \left( C_{1,0,0} \; x + C_{0,1,0} \; y + C_{0,0,1} \; z
\right)
\\
& +  \left( C_{0, 2, 0} \; y^2 + C_{2, 0, 0} \; x^2 + C_{0, 0, 2}
\; z^2
+ C_{1,1,0} \; xy + C_{1,0,1} \; xz + C_{0,1,1} \; yz \right) L + \ldots
\end{split}
\label{eq5.11}
\end{equation}
The leading coefficients $C_{1,0,0} = 0$, $C_{0,1,0} = 1$, $C_{0,0,1} =
0$ are known from the $\lambda \phi^4 / 4$ tree potential.  If we
substitute Eq.\ (\ref{eq5.11}) with these values into Eq.\ (\ref{eq5.10}), we find that the aggregate  term
independent of $L$ is just
\begin{equation}
- 2 \left( C_{0,2,0} \; y^2 + C_{2,0,0} \; x^2 + C_{0,0,2} \; z^2 + C_{1,1,0} \; xy
+ C_{1,0,1} \; xz  + C_{0,1,1} yz \right) + 6y^2 - \frac{3}{2} x^2 = 0,
\label{eq5.12}
\end{equation}
in which case $C_{0,2,0} = 3$, $C_{2,0,0} = -\frac{3}{4}$, and the remaining degree-2
coefficients within Eq.\ (\ref{eq5.10}) are zero:
\begin{equation}
S_{LL} = y + 3y^2 L - \frac{3}{4} x^2 L + \ldots
= \frac{\lambda}{4\pi^2} + \left( \frac{3\lambda^2}{16\pi^4} -
\frac{3g_t^4}{64\pi^4} \right) \log \left( \frac{\phi^2}{\mu^2} \right)
+ \ldots
\label{eq5.13}
\end{equation}
The one-loop ${\cal{O}}(\lambda^2, g_t^4)$ diagrammatic contributions to
the effective potential (\ref{eq5.1}) for RBEWS are easily obtained upon substitution of
Eq.\ (\ref{eq5.13}) into Eq.\ (\ref{eq5.9});  RG-invariance and the one-loop RG-functions
(\ref{eq5.6a})--(\ref{eq5.6d}) are sufficient in themselves to determine ${\cal{O}} (\lambda^2,
g_t^4)$ contributions to the one-loop effective potential (\ref{eq5.1}) obtained
directly from diagrams in Ref.\ \cite{A}.


\section{Summations of Leading Logarithms in RBEWS}

\renewcommand{\theequation}{6.\arabic{equation}}
\setcounter{equation}{0}

The leading-logarithm summation $S_{LL}$ defined by Eq.\ (\ref{eq5.9}) may be
expressed as a series expansion in the Yukawa couplant $x$ [Eq.\ (\ref{eq5.7a})]:
\begin{equation}
S_{LL} = y F_0 (w, \zeta) + \sum_{n=1}^\infty x^n L^{n-1} F_n (w,
\zeta),
\label{eq6.1}
\end{equation}
where we find it convenient to utilize the new variables
\begin{equation}
\zeta \equiv zL, \; \; \; w \equiv 1 - 3yL .
\label{eq6.2}
\end{equation}
The coefficient functions $F_k$ appearing in Eq.\ (\ref{eq6.1}) are the
${\cal{O}}(x^k)$ contributions to the leading logarithm series $S_{LL}$.
Comparison to Eq.\ (\ref{eq5.9}) shows these functions to be themselves
summations over the new variables $\zeta$ and $w$:
\begin{equation}
F_n (w, \zeta) = \sum_{\ell=0}^\infty \sum_{k=0}^\infty C_{n, \ell, k}
\left( \frac{1 - w}{3} \right)^\ell \zeta^k .
\label{eq6.3}
\end{equation}
Upon substituting Eq.\ (\ref{eq6.1}) into the RGE (\ref{eq5.10}), we obtain the following
recursive set of partial differential equations for $F_k (w, \zeta)$:
\begin{gather}
\zeta \left( 1 + \frac{7}{4} \zeta \right) \frac{\partial}{\partial
\zeta} F_0 (w, \zeta) = (1 - w) \left[ w\frac{\partial}{\partial w} +
1 \right] F_0 (w, \zeta),
\label{eq6.4}
\\
\left[ \zeta \left( 1 + \frac{7}{4} \zeta \right)
\frac{\partial}{\partial \zeta} + 2\zeta + (w - 1) w
\frac{\partial}{\partial w} \right] F_1 (w, \zeta) = -\frac{(w-1)^2}{2}
\frac{\partial}{\partial w} F_0 (w, \zeta),
\label{eq6.5}
\\
\begin{split}
&\left[ \zeta \left( 1 + \frac{7}{4} \zeta \right)
\frac{\partial}{\partial \zeta}  +  (w-1) w \frac{\partial}{\partial w} +
(1 + 4\zeta) \right] F_2 (w, \zeta)
\\
&\qquad\qquad\qquad=  \left[ \frac{3}{2} (w-1) \frac{\partial}{\partial w} - \frac{3}{8}
\right] F_1 (w, \zeta) - \frac{3}{4} \left[ (w-1)
\frac{\partial}{\partial w} + 1 \right] F_0 (w, \zeta),
\end{split}
\label{eq6.6}
\\
\begin{split}
&\left[ \zeta \left(1 + \frac{7}{4} \zeta \right) \frac{\partial}{\partial
\zeta} + (w-1) w \frac{\partial}{\partial w} + (k-1 + 2k \zeta)
\right] F_k (w, \zeta)
\\
&\qquad\qquad\qquad= \left[ \frac{3(3k-7)}{8} + \frac{3}{2} (w-1) \frac{\partial}{\partial
w} \right] F_{k-1} (w, \zeta) + \frac{9}{2} \frac{\partial F_{k-
2}}{\partial w} (w, \zeta), \; \; \; (k \geq 3).
\end{split}
\label{eq6.7}
\end{gather}
The functions $F_k (w, \zeta)$ are themselves summations of all leading
logarithms analogous to the functions $S_k [w]$ of Section 2.  Their
successive solutions are obtained below.

\subsection{\bf{\em Solution for $F_0 (w, \zeta)$}}

Consider first Eq.\ (\ref{eq6.4}) when $\zeta$ is fixed at zero:
\begin{equation}
w \frac{d}{dw} F_0 (w, 0) + F_0 (w, 0) = 0
\label{eq6.8}
\end{equation}
We know that the lead term in the series $S_{LL}$ is $y$ [$C_{0,1,0} =
1$ and $C_{1,0,0} = C_{0,0,1} = 0$ in Eq.\ (\ref{eq5.9})], in which case we see
from Eqs.\ (\ref{eq6.1}) and (\ref{eq6.2}) that $F_0 (1,0) = 1$ to ensure that $S_{LL}
\begin{array}{c}{}\\ _{\longrightarrow}\\ ^{L \rightarrow 0} \end{array} y$.
With this initial condition, the solution to Eq.\ (\ref{eq6.8}) is $F(w,0) = 1/w$.
Now suppose we substitute $F_0 (w, \zeta) = G (\zeta) K(w)$ into Eq.\
(\ref{eq6.4}).  This separation of variables implies that
\begin{equation}
\frac{\zeta\left( 1 + \frac{7}{4} \zeta \right) G^\prime
(\zeta)}{G(\zeta)} = (1-w) \left[ \frac{w K^\prime (w)}{K(w)} + 1
\right]
\label{eq6.9}
\end{equation}
We see that when $w = 1$, $G^\prime (\zeta) = 0$, in which case $G$ is
constant.  Since $G(\zeta)$ is independent of $w$ [{\it i.e.}, since the
left-hand side of Eq.\ (\ref{eq6.9}) is independent of $w$], $G$ is constant and
\begin{equation}
F_0 (w, \zeta) = F_0 (w, 0) = 1/w .
\label{eq6.10}
\end{equation}

Indeed this result is not unexpected.  $F_0$ is the sum of all leading
logarithm contributions to the effective potential that do not involve
the Yukawa coupling (i.e. the couplant $x$). Since scalar fields do not
couple directly to $SU(3)$ gluons, the only graphs with external scalar
field legs that contain the strong interaction coupling constant
necessarily contain internal quark lines, which can then emit/absorb
virtual gluons, and such quark lines cannot occur unless a Yukawa
interaction is present to couple scalar field lines to quark-antiquark
pairs.  For the case where the Yukawa couplant $x$ does not occur,
namely, the coefficient $F_0 (w, \zeta)$ in the series $S_{LL}$, there
are no $SU(3)$ fermions coupled to the $F_0$ subset of leading log
graphs, which necessarily implies that $F_0$ is independent of the
$SU(3)$ couplant $z$.  Since $\zeta \equiv zL$ is the source of any $z$
dependence within $F_0 (w,\zeta)$, $F_0 (w, \zeta)$ clearly must be
independent of $\zeta$, as is evident in Eq.\ (\ref{eq6.10}).

\subsection{\bf {\em Solution for $F_1 (w, \zeta)$}}

If we substitute the solution (\ref{eq6.10}) for $F_0 (w, \zeta)$ into Eq.\
(\ref{eq6.5}), we find the right-hand side of this equation is just equal to
$\frac{1}{2} \left[ (w-1) / w \right]^2$.  Since $\left[ (w-1) / w
\right]^2$ is an eigenstate of the operator $(w-1) w (\partial /
\partial w)$ appearing on the left-hand side of Eq.\ (\ref{eq6.5}), one can
choose
\begin{equation}
 F_1 (w, \zeta) = \frac{1}{4} M (\zeta) \left[ (w-1) / w \right]^2
\label{eq6.11}
\end{equation}
and find that $\left[(w-1) / w \right]^2$ factors out of Eq.\ (\ref{eq6.5}) to
yield the first-order ordinary differential equation
\begin{equation}
\left( 1 + \frac{7}{4} \zeta \right) M^\prime (\zeta) + 2
\frac{(\zeta+1)}{\zeta} M(\zeta) = \frac{2}{\zeta}
\label{eq6.12}
\end{equation}
The integrating factor for this equation,
\begin{equation}
g(\zeta) = \exp \left[ \frac{8}{7} \int
\frac{(\zeta+1)}{\zeta(\zeta+4/7)} \; d\zeta \right] = \zeta^2 \left(
\zeta + \frac{4}{7} \right)^{-6/7} ,
\label{eq6.13}
\end{equation}
vanishes at $\zeta=0$.  Thus, the requirement that $M(\zeta)$ not be
singular at $\zeta = 0$ [{\it i.e.}, not be singular in the limit the QCD couplant vanishes]
uniquely specifies the solution
\begin{equation}
M(\zeta)  = \frac{8}{7g(\zeta)} \int_0^\zeta dr \; g(r) r^{-1} (r + 4/7)^{-
1}
 =  \frac{8}{\zeta} + \frac{16}{3\zeta^2} \left[ 1 - \left( 1 +
\frac{7\zeta}{4} \right)^{6/7} \right]
=1-\frac{2}{3}\zeta+\frac{5}{8}\zeta^2+\ldots
\quad .
\label{eq6.14}
\end{equation}
$F_1 (w, \zeta)$ is then obtained by explicit substitution of the final
line of Eq.\ (\ref{eq6.14}) into Eq.\ (\ref{eq6.11}).

\subsection{\bf {\em Solution for $F_2 (w, \zeta)$}}

To obtain a solution to Eq.\ (\ref{eq6.6}), we wish to exploit the identity
\begin{equation}
w(w-1) \frac{\partial}{\partial w} \left[ H(\zeta) \left[ \frac{w-
1}{w}\right]^k\right] = k \; H(\zeta) \left[ \frac{w-1}{w} \right]^k
\label{eq6.15}
\end{equation}
by expressing the $w$-dependence of the right-hand side of Eq.\ (\ref{eq6.6}) as
a series in the form $\sum_{k=0} A_k (\zeta) \left[ (w-1)/w\right]^k$.
From the solution (\ref{eq6.10}) for $F_0$, we see that
\begin{equation}
\left[ (w-1) \frac{\partial}{\partial w} + 1 \right] \frac{1}{w} =
\frac{1}{w^2} = 1 - 2\left( \frac{w-1}{w}\right) + \left(\frac{w-
1}{w}\right)^2.
\label{eq6.16}
\end{equation}
Moreover, we see from Eq.\ (\ref{eq6.11}) that
\begin{equation}
(w-1) \frac{\partial}{\partial w} F_1 (w, \zeta) = \frac{1}{2}
M(\zeta) \left[ \left( \frac{w-1}{w}\right)^2 - \left( \frac{w-1}{w}
\right)^3 \right],
\label{eq6.17}
\end{equation}
in which case the right-hand side of Eq.\ (\ref{eq6.6}) may be expressed as 
\begin{equation}
-\frac{3}{4} + \frac{3}{2} \left( \frac{w-1}{w}\right) + \left[
\frac{21}{32} M(\zeta) - \frac{3}{4} \right] \left( \frac{w-1}{w}
\right)^2 - \frac{3}{4} M(\zeta) \left( \frac{w-1}{w}\right)^3.
\end{equation}  
To solve for $F_2$, we utilize the property (\ref{eq6.15}) within Eq.\ (\ref{eq6.6}) by
writing
\begin{equation}
F_2(w, \zeta) = H_0 (\zeta) + H_1 (\zeta) \left( \frac{w-1}{w}\right)
+ H_2 (\zeta) \left( \frac{w-1}{w}\right)^2 + H_3 (\zeta) \left(
\frac{w-1}{w}\right)^3.
\label{eq6.18}
\end{equation}
We substitute Eq.\ (\ref{eq6.18}) into Eq.\ (\ref{eq6.6}) and equate powers of $\left[ (w-
1)/w\right]^k$ to find that
\begin{gather}
\zeta \left( 1 + \frac{7}{4} \zeta\right) H_0^\prime (\zeta) + (1 +
4\zeta) H_0 (\zeta) = -3/4,
\label{eq6.19}
\\
\zeta \left( 1 + \frac{7}{4} \zeta\right) H_1^\prime (\zeta) + (2 +
4\zeta) H_1 (\zeta) = 3/2,
\label{eq6.20}
\\
\zeta \left( 1 + \frac{7}{4} \zeta\right) H_2^\prime (\zeta) + (3 +
4\zeta) H_2 (\zeta) = \frac{21}{32} M(\zeta) - \frac{3}{4},
\label{eq6.21}
\\
\zeta \left( 1 + \frac{7}{4} \zeta\right) H_3^\prime (\zeta) + (4 +
4\zeta) H_3 (\zeta) = -\frac{3}{4} M(\zeta),
\label{eq6.22}
\end{gather}
where $M(\zeta)$ is given by Eq.\ (\ref{eq6.14}).  The requirement that $\{H_0
(\zeta), H_1 (\zeta),$ $ H_2 (\zeta), H_3 (\zeta) \}$ not be singular
in the $\zeta = 0$ limit in which QCD is turned off uniquely specifies
the following solutions:
\begin{gather}
H_0 (\zeta) = - \frac{3}{7\zeta \left( \zeta +
\frac{4}{7}\right)^{9/7}} \int_0^\zeta \left( r +
\frac{4}{7}\right)^{2/7} dr
= \frac{\left(1 + \frac{7\zeta}{4}\right)^{-9/7} - 1}{3\zeta}
=-\frac{3}{4}+\frac{3}{2}\zeta-\frac{23}{8}\zeta^2+\ldots
,
\label{eq6.23}
\\
H_1 (\zeta) = \frac{6}{7\zeta^2 \left( \zeta +
\frac{4}{7}\right)^{2/7}} \int_0^\zeta dr \; r \left( r +
\frac{4}{7}\right)^{-5/7}
= \frac{2\zeta - 4\left[ 1-\left( 1 + \frac{7\zeta}{4}\right)^{-2/7}\right]}{3\zeta^2}
=\frac{3}{4}-\zeta+\frac{23}{16}\zeta^2+\ldots\quad
,
\label{eq6.24}
\\
\begin{split}
H_2 (\zeta) & =  \frac{8\left( \zeta + 4/7\right)^{5/7}}{7\zeta^3}
\int_0^\zeta dr \frac{r^2}{(r+ 4/7)^{12/7}} \left[ \frac{21}{32} M(r) -
\frac{3}{4} \right]
\\
& =  \frac{1}{\zeta^3} \left[ \frac{20}{3} + \frac{71}{6} \zeta -
\frac{\zeta^2}{3} + \frac{22}{3} \left( 1 + \frac{7\zeta}{4}
\right)^{5/7} - 14 \left( 1 + \frac{7\zeta}{4} \right)^{6/7} \right]
=-\frac{1}{32}-\frac{5}{64}\zeta+\frac{11}{64}\zeta^2+\ldots\quad
,
\end{split}
\label{eq6.25}
\\
\begin{split}
H_3 (\zeta)  = & \frac{4(\zeta + 4/7)^{12/7}}{7\zeta^4}
\int_0^\zeta dr \frac{r^3}{(r+4/7)^{19/7}} \left( -\frac{3 M(r)}{4}
\right)
\\
 = & \frac{1}{\zeta^4} \left[ - \frac{16}{3} - 16\zeta - 12\zeta^2 +
\frac{32}{21} \left( 1 + \frac{7\zeta}{4} \right)^{6/7}
 -   \frac{16}{3} \left( 1 + \frac{7\zeta}{4} \right)^{12/7} +
\frac{64}{7} \left( 1 + \frac{7\zeta}{4} \right)^{13/7} \right]
\\
=&-\frac{3}{16}+\frac{1}{4}\zeta-\frac{61}{192}\zeta^2+\ldots\quad
.
\end{split}
\label{eq6.26}
\end{gather}
Thus $F_2 (w, \zeta)$ is obtained explicitly by substitution of Eqs.\
(\ref{eq6.23}) -- (\ref{eq6.26}) into Eq.\ (\ref{eq6.18}).

\subsection{\bf{\em ${\cal{O}}(x^3)$ Contribution to $S_{LL}$}}

We now consider the $k = 3$ case of Eq.\ (\ref{eq6.7}).  We substitute into this
equation's right-hand side the solutions (\ref{eq6.11}) and (\ref{eq6.18}) for $F_1 (w,
\zeta)$ and $F_2 (w, \zeta)$, and we then expand this side in terms of
powers of $[(w-1)/w]^k$ to obtain the partial differential equation
\begin{equation}
\begin{split}
&\left[ \zeta \left( 1 + \frac{7}{4} \zeta \right) \frac{\partial}{\partial \zeta} + (w-1) w \frac{\partial}{\partial w} 
+ 2 (1 + 3\zeta) \right] F_3 (w, \zeta)
\\
& \qquad=  \frac{3}{4} H_0 (\zeta) + \frac{9}{8} \left( 2H_1 (\zeta) + M(\zeta) \right) \left[ \frac{w-1}{w} \right]
+ \frac{3}{4} \left( 5 H_2 (\zeta) - 2H_1 (\zeta) - 3M (\zeta) \right)
\left[ \frac{w-1}{w} \right]^2
\\
& \qquad\qquad+  \frac{3}{8} \left( 14 H_3 (\zeta) - 8H_2 (\zeta) + 3M(\zeta) \right) \left[ \frac{w-1}{w}\right]^3
- \frac{9}{2} H_3 (\zeta) \left[ \frac{w-1}{w} \right]^4.
\end{split}
\label{eq6.27}
\end{equation}
As before, we express $F_3$ in powers of $\left[ (w-1)/w \right]$,
\begin{equation}
F_3 (w, \zeta) = \sum_{k=0}^4 N_k (\zeta) \left[ \frac{w-1}{w}\right]^k,
\label{eq6.28}
\end{equation}
to obtain the following first-order differential equations for $N_k (\zeta)$:
\begin{gather}
\zeta \left( 1 + \frac{7}{4} \zeta \right) N_0^\prime (\zeta)+2\left( 1
+ 3\zeta\right) N_0 (\zeta) = \frac{3}{4} H_0 (\zeta),
\label{eq6.29}
\\
\zeta \left( 1 + \frac{7}{4} \zeta \right) N_1^\prime (\zeta)+3\left( 1 
+ 2\zeta\right) N_1 (\zeta) = \frac{9}{4} H_1 (\zeta) + \frac{9}{8} M(\zeta),
\label{eq6.30}
\\
\zeta \left( 1 + \frac{7}{4} \zeta \right) N_2^\prime (\zeta)+2\left( 2 
+ 3\zeta\right) N_2 (\zeta) = \frac{3}{4} \left[5H_2 (\zeta) - 2H_1 (\zeta) - 3M(\zeta)\right],
\label{eq6.31}
\\
\zeta \left( 1 + \frac{7}{4} \zeta \right) N_3^\prime (\zeta)+ \left( 5 
+ 6\zeta\right) N_3 (\zeta) = \frac{3}{8} \left[14H_3 (\zeta) - 8H_2 (\zeta) + 3M(\zeta)\right],
\label{eq6.32}
\\
\zeta \left( 1 + \frac{7}{4} \zeta \right) N_4^\prime (\zeta) + 6 (1 +
\zeta) N_4 (\zeta) = - \frac{9}{2} H_3 (\zeta).
\label{eq6.33}
\end{gather}

These can all be solved exactly using the requirement that $N_k
(\zeta)$ is nonsingular at $\zeta = 0$, and their solutions are tabulated
in Appendix C.  However, our analysis of the
effective potential in the next section will prove to be
sensitive only up to ${\cal{O}}(L^4)$ terms in the leading logarithm
series $S_{LL}$ [Eqs.\ (\ref{eq5.9}) or (\ref{eq6.1})].  Thus it is sufficient for us
here to generate series solutions for $F_3$ such that the term $x^3 L^2
F_3 (w, \zeta)$ within $S_{LL}$ is specified up to (and including)
terms of order $L^4$.  Since $\zeta = zL$, we thus need to know $N_0
(\zeta)$ only up to terms quadratic in $\zeta$.  Moreover, since $w - 1 = -3
yL$, we see from Eq.\ (\ref{eq6.28}) that we need to know $N_1 (\zeta)$ only to
terms linear in $\zeta$, and $N_2 (\zeta)$ only to its constant term
$N_2 (0)$.  The functions $N_3 (\zeta)$ and $N_4 (\zeta)$ will not
participate at all in the analysis of the next section, since they
correspond to contributions of order $L^5$ and higher to the series
$S_{LL}$.

From Eqs.\ (\ref{eqC.3})--(\ref{eqC.5}) of Appendix C, we find that
\begin{gather}
N_0 (\zeta) = - \frac{9}{32} + \frac{15}{16} \zeta - \frac{603}{256}
\zeta^2 + \ldots
\label{eq6.39}
\\
N_1 (\zeta) = \frac{15}{16} - \frac{69}{32} \zeta + \ldots
\label{eq6.40}
\\
N_2 (\zeta) = - \frac{447}{512} + {\cal{O}}(\zeta)
\label{eq6.41}
\end{gather}
We substitute these results into Eq.\ (\ref{eq6.28}), and make use of Eq.\ (\ref{eq6.2})
to express $F_3$'s contribution to the leading-logarithm series $S_{LL}$
in terms of the couplants $x$, $y$, and $z$:
\begin{equation}
x^3 L^2 F_3 = \left( - \frac{9x^3}{32} \right) L^2 +
\frac{15x^3}{16} (z - 3y) L^3
+ \left( \frac{x^3}{32} \left[ 207yz - \frac{603}{8} z^2 -
\frac{8343}{16} y^2 \right] \right) L^4 + {\cal{O}}(L^5).
\label{eq6.42}
\end{equation}

\subsection{\bf {\em ${\cal{O}}(x^4)$ and ${\cal{O}}(x^5)$ Contributions to
$S_{LL}$}}

By substituting expressions (\ref{eq6.28}) and (\ref{eq6.18}) for $F_3$ and $F_2$ into
the $k=4$ version of Eq.\ (\ref{eq6.7}), we find that
\begin{equation}
\begin{split}
&\left[ \zeta \left( 1 + \frac{7}{4} \zeta\right)
\frac{\partial}{\partial \zeta} + (3 + 8\zeta) + w (w-1)
\frac{\partial}{\partial w} \right] F_4 (w, \zeta)
\\
 &\qquad\qquad=  \left[ \frac{15}{8} N_0 (\zeta) + \frac{9}{4} H_1 (\zeta)
\right] + \left[ \frac{w-1}{w}\right] \left[ \frac{27}{8} N_1 (\zeta) -
\frac{9}{2} H_1 (\zeta) + \frac{9}{2} H_2 (\zeta)\right]
\\
&\qquad\qquad\qquad +  \left[ \frac{w-1}{w}\right]^2 \left[ \frac{39}{8} N_2 (\zeta) -
\frac{3}{2} N_1 (\zeta) + \frac{9}{4} H_1 (\zeta) - 9H_2 (\zeta) +
\frac{27}{4} H_3 (\zeta)\right]
\\
& \qquad\qquad\qquad +  \left[ \frac{w-1}{w}\right]^3 \left[ \frac{51}{8} N_3 (\zeta) -
3N_2 (\zeta) + \frac{9}{2} H_2 (\zeta) - \frac{27}{2} H_3
(\zeta)\right]
\\
&\qquad\qquad\qquad  +  \left[ \frac{w-1}{w}\right]^4 \left[ \frac{63}{8} N_4 (\zeta) -
\frac{9}{2} N_3 (\zeta) + \frac{27}{4} H_3 (\zeta)\right]
 +  \left[ \frac{w-1}{w}\right]^5 \left[ -6N_4 (\zeta)\right].
\end{split}
\label{eq6.43}
\end{equation}
As before, we define
\begin{equation}
F_4 (w,\zeta) = \sum_{k=0}^5 P_k (\zeta) \left( \frac{w-1}{w}\right)^k
\label{eq6.44}
\end{equation}
so as to generate first order differential equations for $P_0, P_1, \ldots \; , P_5$:
\begin{gather}
\zeta \left( 1 + \frac{7}{4} \zeta \right) P_0^\prime (\zeta) + (3 +
8\zeta) P_0 (\zeta) = \frac{15}{8} N_0 (\zeta) + \frac{9}{4} H_1
(\zeta),
\label{eq6.45}
\\
\zeta \left( 1 + \frac{7}{4} \zeta \right) P_1^\prime (\zeta) + 4
(1 + 2\zeta) P_1 (\zeta)
 = \frac{27}{8} N_1 (\zeta) - \frac{9}{2} H_1 (\zeta) + \frac{9}{2}
H_2 (\zeta),
\label{eq6.46}
\end{gather}
{\it etc}.  In Appendix C we list solutions for $\{N_0 (\zeta), \ldots , N_4 (\zeta)\}$
obtained by solving Eqs.\ (\ref{eq6.29}) -- (\ref{eq6.33}) with the requirement that $N_k (\zeta)$ be
finite at $\zeta = 0$.  Consequently, one could solve differential equations
such as (\ref{eq6.45}) and (\ref{eq6.46}) for $P_k (\zeta)$ by imposing a similar
requirement of finiteness at $\zeta = 0$.  However, as noted above, our
extraction of a scalar-field mass presented in the next section is
sensitive only to terms up to degree-4 in L within the series (\ref{eq6.1}) [or
(\ref{eq5.9})] for $S_{LL}$;  the analysis will be insensitive to terms of
${\cal{O}}(L^5)$.  Since $[(w-1)/w] = -3yL + {\cal{O}}(L^2)$ and $\zeta
= zL$, we see that the only ${\cal{O}}(x^4)$ contributions $x^4 L^3 F_4 (w,
\zeta)$ to $S_{LL}$ relevant to the analysis in the section that
follows arise from knowing $P_0 (\zeta)$ to its term linear in
$\zeta$, and from knowing $P_1 (\zeta)$'s constant lead term $P_1
(0)$.  By substituting the final series in Eq.\ (\ref{eq6.24}) for
$H_1(\zeta)$ and the series (\ref{eq6.39})
for $N_0 (\zeta)$ into Eq.\ (\ref{eq6.45}), we find that
\begin{equation}
P_0 (\zeta) = \frac{99}{256} - \frac{459 \zeta}{512} + {\cal{O}}
(\zeta^2).
\label{eq6.47}
\end{equation}
Similarly, we find from the lead term of the final series contributions within Eqs.\ (\ref{eq6.24}),
(\ref{eq6.25}) and (\ref{eq6.40}) to the right-hand side of Eq.\ (\ref{eq6.46}) that
\begin{equation}
P_1 (\zeta) = -\frac{45}{512} + {\cal{O}} (\zeta) .
\label{eq6.48}
\end{equation}
We can substitute these results into Eq.\ (\ref{eq6.44}) to find the aggregate
${\cal{O}} (x^4)$ contribution to $S_{LL}$ to ${\cal{O}}(L^4)$:
\begin{equation}
x^4 L^3 F_4 (w, \zeta) = \left( \frac{99}{256} x^4 \right)
L^3 + x^4 \left( -\frac{459}{512} z + \frac{135}{512} y \right)
L^4 + {\cal{O}}(L^5).
\label{eq6.49}
\end{equation}

Finally we see that the only contribution to $S_{LL}$ from its
${\cal{O}}(x^5)$ term $x^5 L^4 F_5 (w, \zeta)$ that is degree-4 in 
$L$ is just $x^5 L^4 F_5 (1,0)$.  This constant term is found from 
constant-term contributions to the $k = 5$ version of Eq.\ (\ref{eq6.7}), which 
satisfy the algebraic relation
\begin{equation}
8F_5 (1,0) = 6P_0 (0) + \frac{9}{2} N_1 (0).
\label{eq6.50}
\end{equation}
Using Eqs.\ (\ref{eq6.40}) and (\ref{eq6.47}), we find that $F_5 (1,0) = 837/1024$.  This
result, in conjunction with Eqs.\ (\ref{eq6.42}) and (\ref{eq6.49}), yields all
${\cal{O}}(x^3)$ and higher contributions to $S_{LL}$ that enter into
the extraction of the Higgs mass in the section that follows:
\begin{equation}
\begin{split}
S_{LL}  = & \frac{y}{w} + x \frac{M(\zeta)}{4} \left( \frac{w-
1}{w}\right)^2
+ x^2 L \left[ \sum_{k=0}^3 H_k (\zeta) \left( \frac{w-1}{w}
\right)^k \right]
\\
& +  \left\{ \left( -\frac{9x^3}{32} \right) L^2 + \left(
\frac{15x^3(z-3y)}{16} + \frac{99}{256} x^4 \right) L^3 \right.
\\
&\qquad +  \left[ \frac{x^3}{32} \left( 207yz - \frac{603}{8} z^2 -
\frac{8343}{16} y^2 \right) + \frac{x^4}{512} \left( 135y -
459z \right) 
+ \left. \frac{837}{1024} x^5 \right] L^4 \right\} +
{\cal{O}} (L^5),
\end{split}
\label{eq6.51}
\end{equation}
where $M(\zeta)$ and $H_k (\zeta)$ are given by Eqs.\ (\ref{eq6.14}) and (\ref{eq6.23})
-- (\ref{eq6.26}).

\section{Extraction of $m_\phi$ from the RBEWS Summation-of-Leading-Logarithms Effective Potential}
\renewcommand{\theequation}{7.\arabic{equation}}
\setcounter{equation}{0}
We have seen in the previous section that the RGE (\ref{eq5.10}) can be used to determine all
coefficients of the leading logarithm series $S_{LL}$ for the effective
potential (\ref{eq5.9}).  The result (\ref{eq6.51}) for this series includes summations
of logarithms to all orders in the scalar self-interaction couplant $y$
and QCD gauge-couplant $z$, and to quadratic terms in the $t$-quark
Yukawa couplant $x$.  However, the result (\ref{eq6.51}) also includes
contributions to coefficients in the series (\ref{eq5.9}) of the leading four
powers of $L$ that are valid to {\it all} orders of $\{x,y,z\}$.  Such
terms are sufficient in themselves for a prediction of the scalar-field
mass to all-orders in $\{x,y,z\}$ for the summation-of-leading-logarithms 
series (\ref{eq6.51}).  One can expand this series in powers of
$L = \log (\phi^2/\mu^2)$ and choose $\mu = \langle\phi\rangle$, the vacuum
expectation value of the scalar field:
\begin{equation}
S_{LL}  =  A + B \log (\phi^2 / \langle\phi\rangle^2) + C \log^2 (\phi^2 /
\langle\phi\rangle^2)
 +  D \log^3 (\phi^2 / \langle\phi\rangle^2) + E \log^4 (\phi^2 / \langle\phi\rangle^2) + \ldots ,
\label{eq7.1}
\end{equation}
where, from Eq.\ (\ref{eq6.51}),
\begin{gather}
B = 3y^2 - \frac{3}{4} x^2,
\label{eq7.2}
\\
C = 9y^3 + \frac{9}{4} xy^2 - \frac{9}{4} x^2 y + \frac{3}{2} x^2 z -
\frac{9}{32} x^3,
\label{eq7.3}
\\
D  =  27 y^4 + \frac{27}{2} x y^3 - \frac{3}{2} xy^2 z + 3x^2
yz
 -  \frac{225}{32} x^2y^2 - \frac{23}{8} x^2z^2 + \frac{15}{16} x^3z -
\frac{45}{16} x^3 y + \frac{99}{256} x^4,
\label{eq7.4}
\\
\begin{split}
E  = & 81y^5 + \frac{243}{4} xy^4 - 9xy^3z + \frac{45}{32} x y^2 z^2 -
\frac{69}{16} x^2 yz^2 -  \frac{135}{8} x^2 y^3 + \frac{531}{64} x^2 y^2 z
\\
 &+ \frac{345}{64}
x^2 z^3 - \frac{603}{256} x^3 z^2
 +  \frac{207}{32} x^3 yz - \frac{8343}{512} x^3 y^2 - \frac{459}{512}
x^4 z + \frac{135}{512} x^4 y + \frac{837}{1024} x^5.
\end{split}
\label{eq7.5}
\end{gather}
The coefficient $A$ will include a finite counterterm $K$,
\begin{equation}
A = y + K,
\label{eq7.6}
\end{equation}
so that the definition for the summation-of-leading-logarithms potential,
\begin{equation}
V_{LL} = \pi^2 \phi^4 S_{LL},
\label{eq7.7}
\end{equation}
is inclusive of this counterterm.  We can expand this potential about
$\langle\phi\rangle$ to obtain
\begin{equation}
\begin{split}
V_{LL}  = & \pi^2 \langle\phi\rangle^4 \Biggl[ 
A + (4A + 2B) \left( \frac{\phi-
\langle\phi\rangle}{\langle\phi\rangle}\right)
 +  (6A + 7B + 4C) \left(
 \frac{\phi-\langle\phi\rangle}{\langle\phi\rangle}\right)^2\Biggr.
\\
 &\Biggl.+  \left( 4A + \frac{26}{3} B + 12C + 8D\right) \left( \frac{\phi-
\langle\phi\rangle}{\langle\phi\rangle}\right)^3
 +  \left( A + \frac{25}{6} B + \frac{35}{3} C + 20D +
16E\right) \left( \frac{\phi-\langle\phi\rangle}{\langle\phi\rangle}\right)^4 + \ldots \Biggr].
\end{split}
\label{eq7.8}
\end{equation}
The condition that $\langle\phi\rangle$ is indeed an extremum of this potential
implies that the coefficient of $(\phi - \langle\phi\rangle)/\langle\phi\rangle$ must vanish,
\begin{equation}
A = -B/2,
\label{eq7.9}
\end{equation}
serving to remove entirely the finite counterterm $K$ from the series
(\ref{eq7.8}).  The condition that the $\phi^4$ vertex extracted from the
effective potential coincides with its tree-level value, {\it i.e.}, that 
\begin{equation}
\left. \frac{d^4 V_{LL}}{d \phi^4}\right|_{\langle\phi\rangle}  =
\frac{d^4}{d\phi^4} \left( \frac{\lambda \phi^4}{4} \right) = 24\pi^2 y,
\label{eq7.10}
\end{equation}
implies that the coefficient of $(\phi - \langle\phi\rangle)^4 / \langle\phi\rangle^4$ in the
series (\ref{eq7.8}) is equal to $y$:
\begin{equation}
y = \frac{11}{3} B + \frac{35}{3} C + 20D + 16E.
\label{eq7.11}
\end{equation}
This is a degree-5 equation for $y$, as is evident from Eqs.\ (\ref{eq7.2})--(\ref{eq7.5}),
since the Standard-Model values for the couplants $x$ and $z$ are
known [Eqs.\ (\ref{eq5.7a}) and (\ref{eq5.7c})].  Once $y$ is determined, one can obtain
the scalar-field mass $m_\phi$ explicitly from the coefficient of $(\phi
- \langle\phi\rangle)^2 / \langle\phi\rangle^2$ in Eq.\ (\ref{eq7.8}):
\begin{equation}
\left. m_\phi^2 = \frac{d^2 V}{d\phi^2} \right|_{\langle\phi\rangle} = 8\pi^2
\langle\phi\rangle^2 (B+C).
\label{eq7.12}
\end{equation}
Indeed, the procedure delineated above for extracting $m_\phi$ within
RBEWS is mathematically equivalent to that of Section 4 for 
${\rm M}\!\!\!$/SED. We have employed the series expansion (\ref{eq7.8}) only 
to illustrate the insensitivity of this procedure to terms 
${\cal{O}}(L^5)$ and higher within the series (\ref{eq5.9}) for $S_{LL}$.

In the case at hand, we find three real solutions to the constraint
(\ref{eq7.11}):
\begin{gather}
y_1 = 0.0538,
\label{eq7.13a}
\\
y_2 = -0.278,
\label{eq7.13b}
\\
y_3 = -0.00143,
\label{eq7.13c}
\end{gather}
given the substitution of known numerical values (\ref{eq5.7a}) and (\ref{eq5.7c}) into
$B$, $C$, $D$ and $E$ [Eqs.\ (\ref{eq7.2})--(\ref{eq7.5})].  Only the values (\ref{eq7.13a})
and (\ref{eq7.13b}) correspond to $\langle\phi\rangle$ being a local minimum of the potential;  the
value (\ref{eq7.13c}) yields a negative value for $d^2V / d\phi^2$ at $\langle\phi\rangle$.
If we substitute the value (\ref{eq7.13a}) or the value (\ref{eq7.13b}) into Eqs.\ (\ref{eq7.2})
and (\ref{eq7.3}), and then substitute the resulting values of $B$ and $C$ into
Eq.\ (\ref{eq7.12}), we find corresponding values for the scalar field mass:
\begin{gather}
m_{\phi_1} = \sqrt{8\pi^2 \langle\phi\rangle^2 \left( B(y_1) + C(y_1) \right)} \; \;  = 216
\; {\rm GeV},
\label{eq7.14a}
\\
m_{\phi_2} = \sqrt{8\pi^2 \langle\phi\rangle^2 \left( B(y_2) + C(y_2) \right)} \; \; = 453
\; {\rm GeV}.
\label{eq7.14b}
\end{gather}
In obtaining these values, we have utilized the known value for the
vacuum expectation value $\langle\phi\rangle \; = 246 \; {\rm GeV} \equiv v$ consistent with the
magnitude of the Fermi constant and the $SU(2)$ gauge coupling constant 
$g_2$ characterizing broken electroweak symmetry in the Standard Model.

It is instructive to relate these results to the results of Section 5
for the one-loop effective potential of Coleman and Weinberg.  The
results  $\lambda \cong 3.6$ (i.e., $y = 0.093$), $m_\phi \cong 350 \;
{\rm GeV}$ obtained in that section can be recovered from Eqs.\ (\ref{eq7.9}), (\ref{eq7.11})
and (\ref{eq7.12}) simply by setting $C$, $D$ and $E$ equal to zero while
retaining the value of $B$ in Eq.\ (\ref{eq7.2}) --  {\it i.e.}, by ignoring all terms
subsequent to the term linear in the logarithm in the series (\ref{eq7.1}).  It
is evident that the mass prediction (\ref{eq7.14b}) relies on an unacceptably
large and negative value for the couplant $y$ [Eq.\ (\ref{eq7.13b})].  The
prediction (\ref{eq7.14a}), however, is not only phenomenologically reasonable,
but is also contingent upon a determination (\ref{eq7.13a}) of the scalar 
self-interaction couplant $y$ that is more in line with the known magnitudes
(\ref{eq5.7a}) and (\ref{eq5.7c}) of the couplants $x$ and $z$ than the value $y =
0.093$ following from the purely one-loop treatment of Section 5.
The contributions of $y$ alone to the $\beta$-function (\ref{eq5.8b}) correspond
to the $\beta$-function of an $O(4)$-symmetric scalar field theory,
which has been calculated to four subleading orders in the scalar-field
self-interaction coupling \cite{M}.  Using these results we find that
\begin{equation}
\lim_{\stackrel{_{x \rightarrow 0}}{_{z \rightarrow 0}}}
\mu \frac{dy}{d\mu} = 6y^2 - \frac{39}{2} y^3 + 187.85y^4
- 2698.3y^5 + 47975 y^6 + \ldots \; \; .
\label{eq7.15}
\end{equation}
If $y = 0.093$, the right-hand side of (\ref{eq7.15}) evaluated term by term is
$10^{-2} [5.2 - 1.6 + 1.4 - 1.9 + 3.1 + \ldots]$, whose increasing
magnitudes are indicative of a failure to converge.  However, if $y =
0.0538$, terms on the right-hand side of Eq.\ (\ref{eq7.15}), which are now $10^{-3} [17.5
- 3.04 + 1.58 - 1.22 + 1.16 + \ldots]$, continue to decrease monotonically,
though very slowly.  Similarly, the anomalous dimension of the scalar field 
(\ref{eq5.6d}) is seen from $O(4)$-symmetric scalar field theory to be
proportional to the series \cite{M} 
\begin{equation}
\lim_{\stackrel{_{x \rightarrow 0}}{_{z \rightarrow 0}}}
\gamma \;\; \sim \;\;  y^2 \left[ 1 - \frac{3}{2} y + \frac{195}{16} y^2 - 132.9 y^3 +
\ldots\right].
\label{eq7.16}
\end{equation}
If $y = 0.0930$, the series in square brackets ceases to decrease after
its third term, $[1 - 0.140 + 0.105 - 0.107 + \ldots]$, while the $y =
0.0538$ version of this same series, $[1 - 0.0807 + 0.0353 - 0.0207 +
\ldots]$, continues to decrease. Thus, the aggregate effect of summing 
leading logarithms appears to bring Section 5's 
$m_\phi \cong 350 \; {\rm GeV}$ one-loop estimate, obtained via a
problematical determination of the couplant 
$y$, down to $216 \; {\rm GeV}$ via a determination of $y$ considerably 
closer in magnitude to those of
the known QCD couplant $z$ $(\equiv \alpha_s (\langle\phi\rangle)/\pi)$ and the $t$-quark
Yukawa couplant $x$ \\
$(\equiv g_t^2(\langle\phi\rangle)/4\pi)$. This is demonstrably
more consistent with the convergence of RG-functions when 
subsequent-to-leading logarithms are taken into consideration.

Of course, any future observation of a Higgs boson mass at or near 216
{\rm GeV} is not in itself proof of radiative electroweak symmetry breaking.
This prediction is subject to unknown corrections from 
subsequent-to-leading logarithms within the perturbative series for the
effective potential, though in the next section we present arguments that 
such corrections are likely to be small.  However,
the value of the quartic scalar couplant corresponding to an
${\cal{O}}$(200 {\rm GeV}) Higgs boson is five times larger in RBEWS than in
conventional spontaneous symmetry breaking.  For the latter case, in
which a negative mass term is explicitly present in the tree-level
potential prior to spontaneous symmetry breaking, the scalar field
interaction couplant is predicted to be $y \; (=\lambda / 4\pi^2) =
m_\phi^2 / (8\pi^2 \langle \phi \rangle ^2)$, less than one fifth the value obtained
in Eq.\ (\ref{eq7.13a}) for radiative symmetry breaking if $m_\phi = 216$ {\rm GeV}.
Present indirect standard-model bounds on the Higgs boson mass, which
come from the $\log (m_\phi)$ dependence of $m_t$, $M_W$, $M_{Z^0}$ and
$\Gamma_{Z^0}$ \cite{K}, are insensitive to the quartic scalar-field
self-interaction coupling $\lambda$.  If electroweak symmetry breaking
is indeed radiative, processes such as the $W^+ W^- \rightarrow ZZ$
scattering cross-section which {\it are} sensitive to $\lambda$
\cite{N} should be greatly enhanced relative to standard model
expectations.  In short, if an ${\cal{O}}$(200 {\rm GeV}) Higgs is discovered,
a ``smoking gun'' indication of symmetry breaking along the lines
proposed in this work would be a factor of 30 enhancement of
$\sigma(W^+_L W^-_L \rightarrow Z_L^0 Z_L^0)$ relative to conventional
standard-model expectations.


\section{Discussion}

\renewcommand{\theequation}{8.\arabic{equation}}
\setcounter{equation}{0}

\subsection{\bf {\em Perturbative Consistency and Residual Scale Dependence}}

The result $m_\phi = 216 \; {\rm GeV}$ is not contingent on any fine-tuning of
the known-interaction couplants $x$ and $z$.  Indeed, if we let $x$ and
$z$ go to zero, the scalar field mass increases only slightly from $216
\; {\rm GeV}$ to $221 \; {\rm GeV}$, with the value for $y$ determined via Eq.\
(\ref{eq7.11}) correspondingly increasing from 0.0538 to 0.0541.  At this
juncture, however, we cannot know if next-to-leading-logarithm
contributions to the effective potential serve (or fail) to destabilize
this summation-of-leading-logarithms solution.  The effect of summing such
nonleading-logarithm contributions is clearly an area for further
investigation.  What is encouraging, however, is the fact that the value
of $y$ obtained from summing leading logarithms appears to be not very
different than the known magnitudes of the dominant contributing
couplants $x$ and $z$.  Logarithms subsequent to leading are accompanied
by additional powers of the couplants $\{x,y,z\}$.  Thus, 
next-to-leading-logarithm contributions to $B$, $C$, $D$ and $E$ are
respectively degree 3, 4, 5 and 6 in the couplants $\{x,y,z\}$, one
degree higher than the leading-logarithm contributions listed in Eqs.\
(\ref{eq7.2})--(\ref{eq7.5}).  If $\{x,y,z\}$ are all comparably ``small,'' one might
expect such subsequent contributions to be perturbatively suppressed.

Such perturbative consistency is supported by an examination of the
residual renormalization-scale dependence of the leading logarithm
effective potential.  To see this, we first compare the residual scale
dependence occurring ...

\begin{figure}[htb]
\centering
\includegraphics[scale=0.6]{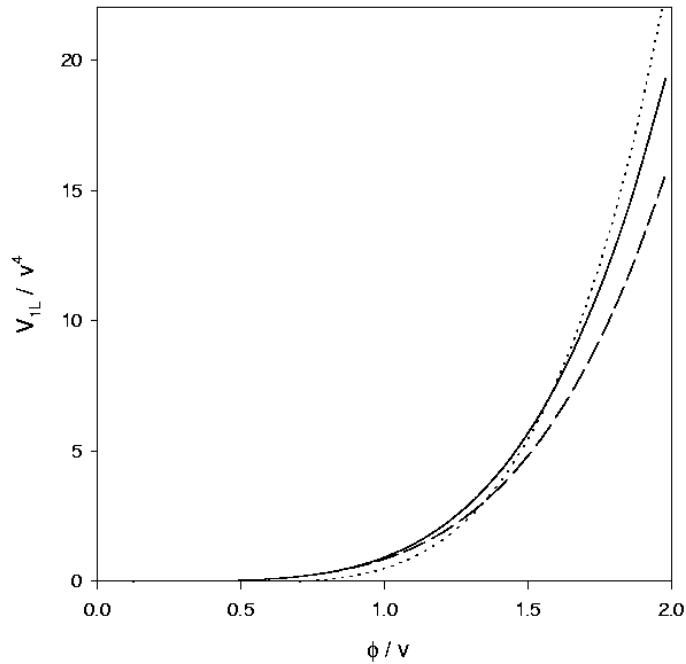}
\caption{
The residual renormalization-scale dependence of Eq.\
(\protect\ref{eq8.1}), the leading-logarithm contribution to the one-loop effective
potential, with couplant and field values evolving from $\mu = v$ as
indicated in the text.  
The top, middle and bottom curves at the right boundary of the figure correspond
respectively to $\mu = 2v$, $\mu = v$ and $\mu = v/2$.
}
\label{fig4}
\end{figure}

\begin{figure}[htb]
\centering
\includegraphics[scale=0.6]{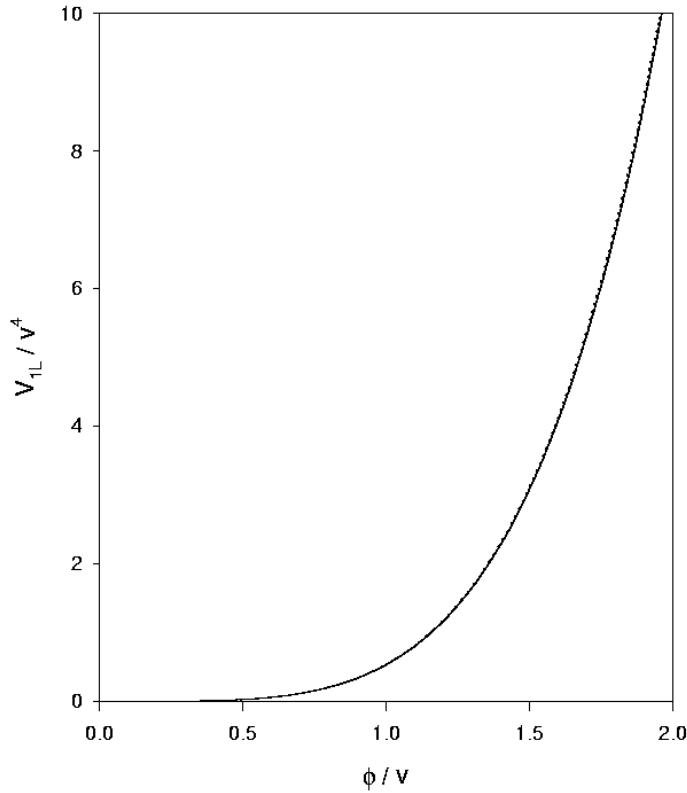}
\caption{
The residual renormalization-scale dependence of Eq.\
(\protect\ref{eq8.2}), the summation-of-leading-logarithms potential truncated after
${\cal{O}}(x^2)$.  Evolution of couplants and $\phi(\mu)$ is as in
Fig.\ \protect\ref{fig4}.  The dotted and solid curves (which overlap almost completely) correspond respectively to $\mu = v/2$ and
$\mu = 2v$;  the $\mu = v$ curve, which falls between these two, has
been omitted for visual clarity.
}
\label{fig5}
\end{figure}

\begin{itemize}

\item[1)]...within
\begin{equation}
V_{1L} = \pi^2 \phi^2 (\mu) \left[ y(\mu) + \left( 3y^2(\mu) - \frac{3}{4} x^2
(\mu) \right) \log \left( \frac{\phi^2 (\mu)}{\mu^2} \right) \right],
\label{eq8.1}
\end{equation}
the leading-logarithm contribution to the {\it one-loop} potential [Fig. \ref{fig4}],
and ...
\item[2)]...within 
\begin{equation}
V_{LL}^{(2)} = \pi^2 \phi^2 (\mu) \left[ y(\mu) + x(\mu) F_1 \left[ w(\mu), \zeta(\mu) \right] 
+  x^2 (\mu) L(\mu) F_2 \left[ w(\mu), \zeta(\mu) \right] \right],
\label{eq8.2}
\end{equation}
the summation-of-leading logarithms series [Fig. \ref{fig5}] truncated after ${\cal{O}}(x^2)$, where
\begin{equation}
L(\mu) = \log \left( \frac{\phi^2 (\mu)}{\mu^2}\right), \; \; w(\mu) = 1 - 3y (\mu) L(\mu),
\; \; \zeta (\mu) = z(\mu) L(\mu).
\label{eq8.3}
\end{equation}
\end{itemize}
Eq.\ (\ref{eq8.1}) is just the one-loop potential (\ref{eq5.1}) without its finite $K
\phi^4$ counterterm, which (since it is more than one degree higher in
couplants than in the logarithm) ultimately generates summations of
subsequent-to-leading logarithms upon incorporation of 
two-and-higher-loop order RG-functions within the RG equation (\ref{eq5.5}).  
The finite counterterm is similarly excluded from Eq.\ (\ref{eq8.2}).

In Figs.\ \ref{fig4} and \ref{fig5}, $x(\mu)$, $y(\mu)$, $z(\mu)$ and $\phi (\mu)$ evolve
from initial values at $\mu = v$ via the one-loop RG-functions (\ref{eq5.8a})--(\ref{eq5.8d}).  
For the three couplants, these initial values are $x(v) = 0.0253$, 
$y(v) = 0.0538$, and $z(v) = 0.0329$ [Eqs.\ (\ref{eq5.7a}), (\ref{eq5.7c}) and (\ref{eq7.13a})].
The field's initial value $\phi(v)$ is an input parameter exhibited
along the abscissae of both figures in units of the vacuum expectation
value $v$.  We see from Figure \ref{fig4} that $V_{1L}$ varies substantially from
$\mu = v/2$ to $\mu=2v$, but that this variation all but vanishes
(Figure \ref{fig5}) when leading logarithms are summed [Eq.\ (\ref{eq8.2})].

\begin{figure}[htb]
\centering
\includegraphics[scale=0.6]{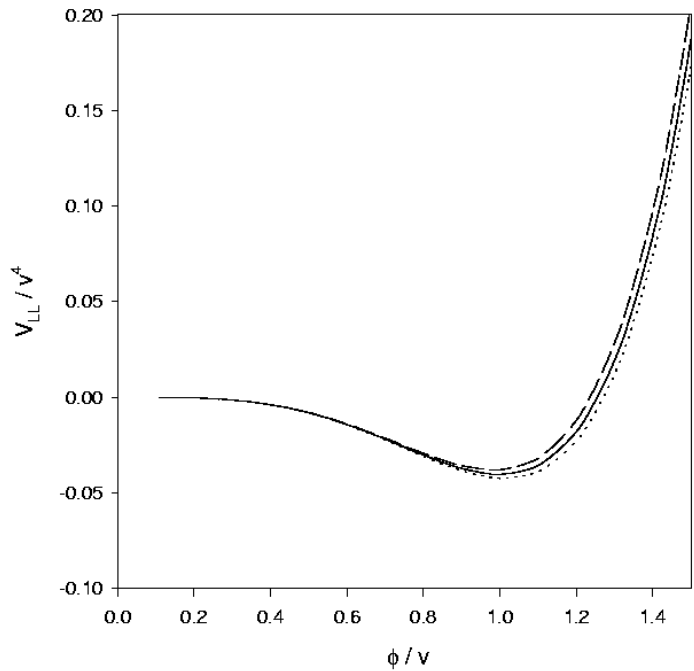}
\caption{
The residual renormalization-scale dependence of the
effective potential obtained through augmentation of Eq.\ (\protect\ref{eq8.2}) with its
appropriate $K\pi^2 \phi^4$ counterterm [Eq.\ (\protect\ref{eq8.4})].  As discussed in
the text, renormalization scale-dependence of this counterterm (which is
a sum of non-leading logarithm terms) is necessarily next-order in the
RGE;  consequently, this counterterm contribution has been assumed to be
RG-invariant.  The evolution of couplants and $\phi(\mu)$ from their
values at $\mu = v$ is as described in the text, and corresponds to that
of Fig.\ \protect\ref{fig7}.  
 As in
Figs.\ \ref{fig4} and \ref{fig5}, the horizontal axis is $\phi(v)/v$, and the vertical axis is
the corresponding value of $V^{LL}_{eff}/v^4$ for the each curve's choice of $\mu$.
The top, middle and bottom curves respectively
correspond to $\mu = v/2$ ($m_\phi = 208$ {\rm GeV}), $\mu = v$ ($m_\phi = 216
{\rm GeV}$), and $\mu = 2v$ ($m_\phi = 217$ {\rm GeV}).
}
\label{fig6}
\end{figure}

Indeed such diminution of residual scale dependence through summation of
logarithms is observed within RG improvement of a wide spectrum of
perturbative calculations \cite{C}. However, if we assume
such residual scale dependence to be indicative of next-order
corrections, we can then expect only modest departures from the $m_\phi
= 216$ {\rm GeV} prediction obtained at $\mu = v$.  The counterterm $K[x(\mu),
y(\mu), z(\mu)] \phi^4$ is necessarily more than degree-two in
couplants, and is degree zero in the logarithm $L$.  Consequently this
term can be partitioned into terms that contribute to the summation of
successively {\it subleading} logarithms, as leading logarithm terms
(\ref{eq5.9}) are only one degree lower in $L$ than in aggregate powers of the
couplants $x$, $y$ and $z$.  Thus the counterterm $K \phi^4$ does not
enter subsequent terms via the leading-logarithm RGE (\ref{eq5.10}).
In 
Figure \ref{fig6}, we evaluate the summation-of-leading-logarithms effective potential
(\ref{eq8.2}) augmented by the $K\phi^4$ counterterm, which was obtained above
by requiring that the coefficient of $(\phi \; - \langle \phi \rangle)^4 / \langle \phi \rangle ^4$
in Eq.\ (\ref{eq7.8}) equal $y$:
\begin{equation}
K = - \left( \frac{25}{6} B + \frac{35}{3} C + 20D + 16E \right),
\label{eq8.4}
\end{equation}
for $\{B, C, D, E \}$ as given by Eqs.\ (\ref{eq7.2}) -- (\ref{eq7.5}) with $x = 0.0253$, $y =
0.0538$ and $z = 0.0329$, as obtained earlier.  Since $K \phi^4$ is not a
term contributing to the sum of leading logarithms within the original
perturbative series (\ref{eq5.9}), we assume this term to be an RG-invariant
contribution (in the leading-log sense) to the effective potential,
whose residual $\mu$-dependence is assumed to reside entirely in the
contribution (\ref{eq8.2}) from the summation of leading logarithms.  We see
from Figure \ref{fig6} that this construction of the effective potential varies
controllably over the range $v/2 \leq \mu \leq 2v$.\footnote{A similar
construction based upon the one-loop contribution (\ref{eq8.1}) exhibits
much larger variation in $m_\phi$ over this same range of $\mu$.}
Moreover, we find over this same range of $\mu$ that the value of
$m_\phi$ extracted from this potential varies only from $208$ {\rm GeV} at
$\mu = v/2$ to $217$ {\rm GeV} at $\mu = 2v$.  If we assume such scale
dependence to be indicative of subsequent subleading-logarithm
corrections, we then can expect only modest departures from the $m_\phi
= 216$ {\rm GeV} prediction at $\mu = v$.  Such scale uncertainties in
$m_\phi$ dominate any uncertainties in $m_\phi$ deriving from the
couplant values themselves ({\it i.e.}, the error in $g_t (v)$ and $\alpha_s
(M_z)$), which affect $m_\phi$ only negligibly.

\subsection{\bf {\em Large-Field Behaviour of the Effective Potential}}

The leading logarithm contribution to the effective potential is given
by the series (\ref{eq6.1}), with $F_0$ given explicitly by (\ref{eq6.10}), $F_1$ given
by Eqs.\ (\ref{eq6.11}) and (\ref{eq6.14}), $F_2$ given by Eqs.\ (\ref{eq6.18}) and (\ref{eq6.23})--(\ref{eq6.26}), and
$F_3$ given by Eq.\ (\ref{eq6.28}) and Eqs.\ (C.4) -- (C.7) of Appendix C.  The solutions
to Eqs.\ (\ref{eq6.5}) -- (\ref{eq6.7}) for $F_n (w, \zeta)$ are all of the general form
\begin{equation}
F_n (w, \zeta) = \sum_{k=0}^{n+1} f_{n,k} (\zeta) \left[ \frac{w-1}{w}
\right]^k.
\label{eq8.5}
\end{equation}
In this notation, we see from Sections 6.2, 6.3 and 6.4 that $M(\zeta)/4
= f_{1,2} \; (f_{1,0} = f_{1,1} = 0)$, $\{H_0, H_1, H_2, H_3\} = \{f_{2,0},
f_{2,1}, f_{2,2}, f_{2,3} \}$, and $\{N_0, N_1, N_2, N_3, N_4 \} =
\{f_{3,0}, f_{3,1}, f_{3,2}, f_{3,3}, f_{3,4} \}$.  If we substitute Eq.\
(\ref{eq8.5}) into Eq.\ (\ref{eq6.7}) and make use of the identities
\begin{gather}
(w-1) w \frac{d}{dw} \left[ \frac{w-1}{w} \right]^k = k \left[ \frac{w-
1}{w} \right]^k,
\label{eq8.6a}
\\
(w-1) \frac{d}{dw} \left[ \frac{w-1}{w} \right]^k = k \left[ \frac{w-
1}{w} \right]^k - k \left[ \frac{w-1}{w} \right]^{k+1},
\label{eq8.6b}
\\
\frac{d}{dw} \left[ \frac{w-1}{w} \right]^k = k \left( \left[ \frac{w-
1}{w} \right]^{k-1} - 2 \left[ \frac{w-1}{w} \right]^{k} + \left[ \frac{w-1}{w} \right]^{k+1} \right),
\label{eq8.6c}
\end{gather}
we then obtain the following recursion relation for $f_{p,k} (\zeta)$
when $p \geq 3$:
\begin{equation}
\begin{split}
& \left[ \frac{7\zeta^2}{2} \frac{d}{d\zeta} + 4p\zeta \right] f_{p,k}
  + \left[ 2\zeta \frac{d}{d\zeta} + 2(p-1) + 2k \right] f_{p,k}
\\
&\qquad =  \left[ \frac{9p-21}{4} + 3k \right] f_{p-1, k} - 3 (k-1) f_{p-1, k-1}
 +  \frac{9}{2} \left[ (k-1) f_{p-2, k-1} - 2k f_{p-2, k} + (k+1) f_{p-2, k+1} \right],
\end{split}
\label{eq8.7}
\end{equation}
where $f_{p,k} \equiv 0$ when $k < 0$ or $k > p+1$, and where $f_{p,k}$
is analytic (finite) at $\zeta = 0$.

\begin{figure}[htb]
\centering
\includegraphics[scale=0.6]{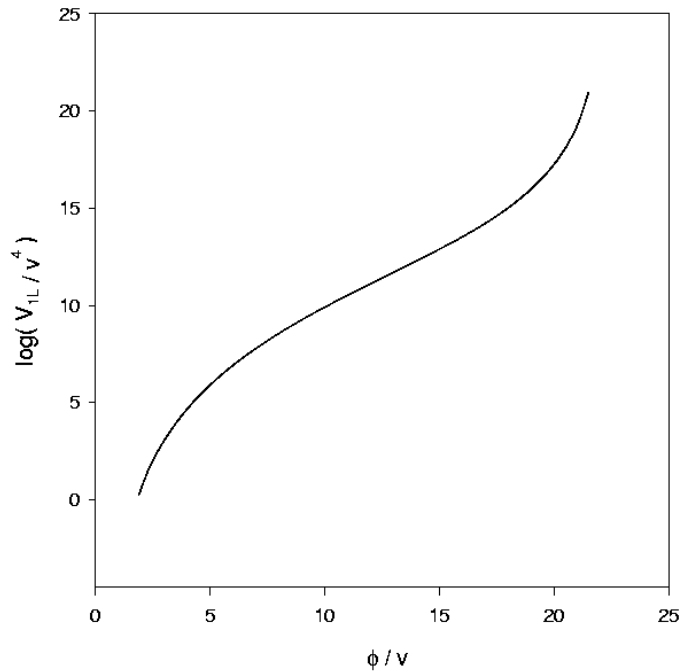}
\caption{
The continuation of the $\mu = v$ curve of Figure \protect\ref{fig6} to
large values of $\phi$.  The ordinate is now logarithmic, as indicated.
}
\label{fig7}

\end{figure}

One of the motivations for summing leading logarithms is to ascertain
the large logarithm behaviour of the effective potential, behaviour
corresponding to the potential in either the large-field or zero-field
limit.  For the large-field case, one is not able to extrapolate past
the $w = 0$ poles characterizing every $F_n (w, \zeta)$ in the series
(\ref{eq6.1}) [as evident from Eqs.\ (\ref{eq6.10}), (\ref{eq6.11}), (\ref{eq6.18}), (\ref{eq6.28}) and,
generally speaking, (\ref{eq8.5})].  Of course, $w = 0$ corresponds to a Landau
pole at $L = 1/3y$ [Eq.\ (\ref{eq6.2})], which implies singular behaviour of the
summation-of-leading logarithms effective potential at $\phi = v \exp
[1/6y] \cong 22.2v$ for our $y = 0.0538$ solution (\ref{eq7.13a}).  In Fig.\ \ref{fig7} we
plot the effective potential of Figure \ref{fig6} for $\mu = v$ over the range
$1.5 v < \phi < e^{1/6y} v$ to illustrate the singularity occurring at
the latter bound.  Such summation-of-logarithms singularities occur in
other contexts and are not necessarily singularities of the
function itself represented by the series. Artificial
singularities in the series expression for the RG-invariant couplant are
discussed in Ref.\ \cite{G};  similar singularities in the perturbative
electron-positron annihilation cross-section are discussed in
refs. \cite{D,P}. Moreover, the terms $S_0$, $S_1$, $S_2$ and $S_3$ within 
the series for the effective potential of ${\rm M}\!\!\!$/SED, as considered 
in Section 3, also exhibit such an ultraviolet singularity at 
$1 - \frac{5}{12} y L = 0$, where $y$ is the scalar couplant of that theory.

Since $w = 1 - 3y \log (\phi^2 / v^2)$, we see that as $\phi$ increases
from its vacuum expectation value $v$ to $v \exp \left[ 1/6y \right]$, $w$
approaches zero from above.  We can show from Eqs.\ (\ref{eq8.5}) and (\ref{eq8.7}),
however, that as $w$ approaches zero from above, all $F_n$ diverge {\it
positively};  {\it i.e.}
\begin{equation}
\lim_{w \rightarrow 0^+} F_n (w, \zeta) \rightarrow + \infty,
\label{eq8.8}
\end{equation}
consistent with each term of the series $S_{LL}$ (\ref{eq6.1}) within $V_{LL} = \pi^2 \phi^4
(S_{LL} + K)$ being bounded {\it from below} prior to the singularity.
To see this, first note from Eq.\ (\ref{eq8.5}) that as $w \rightarrow 0$, the
asymptotic behaviour of each $F_n (w, \zeta)$ is dominated by the
coefficient $f_{n, n+1}$:
\begin{equation}
F_n (w, \zeta)
\begin{array}{c}{}\\_{\longrightarrow} \\ ^{w \rightarrow
0^+}
\end{array}
(-1)^{n+1} f_{n, n+1}
(\zeta) / w^{n+1}.
\label{eq8.9}
\end{equation}
Note also that $\zeta$ remains small in this limit:  when $w \rightarrow
0^+$, $L \rightarrow 1/3y$ and $\zeta \rightarrow z/3y = 0.204$ [$z (v) = 0.0329$ and $y
(v) = 0.0538$].  But if $\zeta$ is small, we see from Eq.\ (\ref{eq8.7}) that
leading contributions to the power series for $f_{n, n+1} (\zeta)$
satisfy the recursion relation \footnote{Since $f_{p,k} = 0$ when $k > p
+ 1$, terms $f_{n-1, n+1}$, $f_{n-2, n}$, etc. appearing in Eq.\ (\ref{eq8.7})
all vanish when $p = n$, $k = n+1$.}
\begin{equation}
f_{n, n+1} (0) = - \frac{3}{4} f_{n-1, n} (0).
\label{eq8.10}
\end{equation}
Since $f_{3,4} (0) = N_4 (0) = 9/64$ [Eq.\ (C.7)], and since $\zeta$ is
small, we see from Eqs.\ (\ref{eq8.9}) and (\ref{eq8.10}) that
\begin{equation}
F_n (w, \zeta)
\begin{array}{c}{}\\_{\longrightarrow} \\ ^{w \rightarrow
0}
\end{array}
(3^{n-1}/4^n) w^{-(n+1)}, \; \; \; n \geq 3.
\label{eq8.11}
\end{equation}
Thus, if $n \geq 3$, $F_n \rightarrow + \infty$ as $w$ approaches zero
from above, {\it i.e.}, as the field $\phi$ increases to approach the
singularity at $L = 1/3y$ from below.  This behaviour also characterizes
$F_n$ for $n < 3$,
\begin{equation}
F_0 = 1/w, \;\; F_1 \begin{array}{c}{}\\_{\longrightarrow} \\ ^{w \rightarrow
0}
\end{array}
M(0) / 4w^2 = 1/4w^2, \; \;
F_2 \begin{array}{c}{}\\ _{\longrightarrow} \\ ^{w \rightarrow
0}
\end{array}
-H_3 (0) / w^3 = 3/16w^3,
\label{eq8.12}
\end{equation}
as is evident from Eqs.\ (\ref{eq6.14}) and (\ref{eq6.26}).  Consequently, we see that
incorporation of arbitrarily many terms of the series (\ref{eq6.1}) into the
effective potential necessarily leads to a potential that diverges
positively as $\phi$ increases to approach the singularity at $w = 0$,
corresponding to an effective potential which is bounded from below
in the $w > 0$ region where the series (\ref{eq6.1}) is meaningful.
Moreover, it should be noted from the large $y$ limit of Eq.\  (\ref{eq5.8b}) for the evolution of $y(\mu)$ that
$y(\mu)=y(v)/\left[1-6y(v)\log(\mu/v)\right]$, an expression which exhibits a Landau pole at $\mu=v\exp{[1/6y(v)]}$ in correspondence
with the singularity discussed above.  Consequently, one may argue (as in ref.\ \cite{quiros}) that the scale $\Lambda$ for new physics
must occur prior to this Landau pole, in which case $\Lambda\le 5.5\,{\rm TeV}$ within the context of {\em radiative} (as opposed to
conventional) spontaneous symmetry breaking.

\subsection{\bf {\em Method of Characteristics and Electroweak-Couplant Corrections}}
An alternative RG approach to incorporating higher order
effects is the method of characteristics \cite{F,Y,Z}, as opposed to use
of the explicit forms (\ref{eq2.10}) and (\ref{eq5.9}) for the leading logarithm
contributions to the effective potential.  Both approaches are shown in
Ref.\ \cite{AA} to lead to equivalent results, including order-by-order
singularities.  However, the method of characteristics has the
advantage of not requiring secondary expansions, such as Eqs.\ (\ref{eq2.21}) or
(\ref{eq6.1}).
The closed-form solution to the RG equation (\ref{eq5.10}) obtained by the method of characteristics
is equivalent to the summation of leading logarithms \cite{A,E,Z} and is given by
\begin{equation}
V_{eff}^{LL}=\pi^2\bar y\left({L}/{2}\right)\bar\phi^4\left({L}/{2}\right)=
\pi^2\bar y\left({L}/{2}\right)\phi^4
\exp{\left[-3\int_0^{L/2}\bar x(t)\,dt\right]},
\label{V_eff_char}
\end{equation}
where $\bar x(t)$, $\bar y(t)$, $\bar z(t)$, and $\bar \phi(t)$ are characteristic functions defined by the differential equations
and initial conditions
\begin{gather}
\frac{d\bar z}{dt}=-\frac{7}{2}\bar z^2,~\bar z(0)=z;
\label{char_z}
\\
\frac{d\bar x}{dt}=\frac{9}{4}\bar x^2-4\bar x\bar z,~\bar x(0)=x;
\label{char_x}
\\
\frac{d\bar \phi}{dt}=-\frac{3}{4}\bar x\bar\phi,~\bar \phi(0)=\phi;
\label{char_phi}
\\
\frac{d\bar y}{dt}=6\bar y^2+3\bar x\bar y-\frac{3}{2}\bar x^2,~\bar y(0)=y,
\label{char_y}
\end{gather}
in obvious correspondence with the one-loop RG functions defined by Eqs.\ (\ref{eq5.8a})--(\ref{eq5.8d}). These differential
equations lead to series solutions
\begin{gather}
\begin{split}
\bar{\phi}\left( t\right) &=\phi \exp \left[ -\frac{3}{4}\int_{0}^{t}%
\bar{x}\left( s\right) ds\right] =\phi \Biggl[ 1-\frac{3}{4}xt+\left( -%
\frac{9}{16}x^{2}+\frac{3}{2}xz\right) t^{2}+\left( -\frac{45}{64}x^{3}+%
\frac{9}{4}x^{2}z-\frac{15}{4}xz^{2}\right) t^{3}\Biggr.
\\
&\qquad \Biggl. +\left( -\frac{135}{128}x^{4}+\frac{135}{12}x^{3}z-\frac{63}{8}%
x^{2}z^{2}+\frac{165}{16}xz^{3}\right) t^{4}+\cdots \Biggr]
\end{split}
\label{char_phi_sol}
\\
\begin{split}
\bar{y}\left( t\right) =&y+\left( 6y^{2}-\frac{3}{2}x^{2}+3xy\right)
t+\left( 36y^{3}-\frac{9}{8}yx^{2}+27xy^{2}-\frac{45}{8}x^{3}-6xyz+6x^{2}z%
\right) t^{2} \\
&+\left( 216y^{4}+18y^{2}x^{2}+216y^{3}x-\frac{387}{32}x^{4}-\frac{477}{16}%
yx^{3}-48xzy^{2}-\frac{15}{2}yx^{2}z+\frac{69}{2}%
x^{3}z+15xyz^{2}-23x^{2}z^{2}\right) t^{3} \\
&+\Biggl( 1296y^{5}+\frac{165}{4}yx^{2}z^{2}+\frac{891}{8}yx^{3}z-360y^{3}xz+%
\frac{675}{2}x^{2}y^{3}+1620xy^{4}+\frac{345}{4}z^{3}x^{2}-\frac{513}{12}%
x^{5}\Biggr. \\
&\qquad\Biggl. +\frac{3501}{32}x^{4}z-\frac{2643}{16}x^{3}z^{2}+\frac{225}{2}%
xz^{2}y^{2}-\frac{165}{4}yxz^{3}-\frac{22059}{256}yx^{4}-\frac{585}{4}%
x^{2}y^{2}z-\frac{7695}{32}x^{3}y^{2}\Biggr) t^{4}+\cdots~~.
\end{split}
\label{char_y_sol}
\end{gather}
which, when substituted into the intermediate expression in Eq.\ (\ref{V_eff_char}), lead to recovery of the series coefficients
$\{B,C,D,E\}$ [Eqs.\ (\ref{eq7.2})--(\ref{eq7.5})].
Indeed, the potential (\ref{V_eff_char}), as constructed from one-loop RG equations
(\ref{char_z}) -- (\ref{char_y}), provides the easiest means of estimating the effect of the subdominant
electroweak gauge couplants $r\equiv g_2^2/4\pi^2$ and $s\equiv g'^2/4\pi^2$ on the Section 7 predictions for
radiatively broken electroweak symmetry.  These couplants, which evolve from initial values
$g_2^2(v)/4\pi^2=\bar r(0)=0.0109$, $g'^2(v)/4\pi^2=\bar s(0)=0.00324$
[consistent with values (\ref{eq5.3a}) and (\ref{eq5.3b}) for $g_2(M_Z)$ and $g'(M_Z)$] via the one-loop RG equations
\cite{ford}
\begin{equation}
\frac{d\bar r}{dt}=-\frac{19}{12}\bar r^2~,\quad \frac{d\bar s}{dt}=\frac{41}{12}\bar s^2
\label{run_gauge}
\end{equation}
are seen to enter the potential (\ref{V_eff_char}) via the following additional electroweak (ew) contributions \cite{I}
to the right-hand sides of Eqs.\ (\ref{char_x})-- (\ref{char_phi}):
{\allowdisplaybreaks
\begin{gather}
\Delta_{ew}\left(\frac{d\bar x}{dt}\right)=-\frac{9}{8}\bar x\bar r-\frac{17}{24}\bar x\bar s,
\label{ew_x}\\
\Delta_{ew}\left(\frac{d\bar \phi}{dt}\right)=-\bar\phi\left[-\frac{9}{16}\bar r-\frac{3}{16}\bar s\right],
\label{ew_phi}\\
\Delta_{ew}\left(\frac{d\bar y}{dt}\right)=-\frac{9}{4}\bar y \bar r-\frac{3}{4}\bar y\bar s+\frac{3}{32}\bar s^2
+\frac{3}{16}\bar r\bar s+\frac{9}{32}\bar r^2
\label{ew_y}
\end{gather}
}
The above corrections allow incorporation of the contributions from the electroweak couplants $r$ and $s$ into the
Eq.\ (\ref{eq7.1}) series
coefficients $\{B, C, D, E\}$:
{\allowdisplaybreaks
\begin{gather}
\Delta_{ew}B=\frac{9}{64}r^2+\frac{3}{32}rs+\frac{3}{64}s^2
\label{ew_B}
\\
\begin{split}
\Delta_{ew}C&=-\frac{33}{1024}r^3+r^2\left(\frac{43}{1024}s+\frac{27}{64}y-\frac{27}{256}x\right)
+r\left(-\frac{27}{16}y^2+s\left(-\frac{9}{128}x+\frac{9}{32}y\right)+\frac{127}{1024}s^2\right)
\\
&+\frac{91}{1024}s^3+s^2 \left(-\frac{9}{256}x+\frac{9}{64}y\right)+s\left(-\frac{9}{16}y^2+\frac{1}{8}x^2\right)
\end{split}
\label{ew_C}\\
\begin{split}
\Delta_{ew}D=&
{\frac {17}{512}}{{ r}}^{4}+ \left( -{\frac {195}{512}}{
 y}+{\frac {587}{36864}}{ s}+{\frac {213}{4096}}{ x
} \right) {{ r}}^{3}
\\&
+ \left( {\frac {177}{64}}{{ y}}^{2}+{
\frac {9}{128}}{ x}{ z}+ \left( -{\frac {119}{512}}{
 y}-{\frac {67}{4096}}{ x} \right) { s}+{\frac {27}{128
}}{ x}{ y}-{\frac {405}{2048}}{{ x}}^{2}+{\frac {
2393}{36864}}{{ s}}^{2} \right) {{ r}}^{2}
\\
&+\left[-{\frac {81}{8}}{{ y}}^{3}
+{\frac {27}{16
}}{ y}{{ x}}^{2}
+{\frac {
27}{512}}{{ x}}^{3}-{\frac {135}{64}}{ x}{{ y}}^{2}
 + \left( {
\frac {99}{64}}{{ y}}^{2}+{\frac {3}{64}}{ x}{ z}-{
\frac {135}{1024}}{{ x}}^{2}+{\frac {9}{64}}{ x}{ y
} \right) { s}
\right.
\\&
\quad\quad\left.
+ \left( -{\frac {447}{4096}}{ x}+{\frac {37}{512}}{ y}
 \right) {{ s}}^{2}
+{\frac {7331}{36864}}{{ s}}^{3}
 \right] { r}
 \\
 &
 + \left( -{\frac {
27}{8}}{{ y}}^{3}-{\frac {53}{64}}{ x}{{ y}}^{2}
+\frac{13}{16}{ y}{{ x}}^{2}+{\frac {49}{512}}{{ x}}^{3}-{\frac
{1}{3}}{{ x}}^{2}{ z} \right) { s}
 + \left( {\frac {5}{16}}{{ y}}^{2}-{\frac {
53}{6144}}{{ x}}^{2}+{\frac {9}{128}}{ x}{ y}
+{
\frac {3}{128}}{ x}{ z} \right) {{ s}}^{2}
\\&
+ \left( {\frac {73}{512}}{ y}-{
\frac {347}{4096}}{ x} \right) {{ s}}^{3}
+{\frac {2029}{12288}}{{ s}}^{4}
\end{split}
\label{ew_D}
\\
\begin{split}
\Delta_{ew}E=&
   -{\frac {11929}{393216}}{{ r}}^{5}+ \left( -{\frac {22367}{
1179648}}{ s}+{\frac {6891}{16384}}{ y}-{\frac {7497}{
262144}}{ x} \right) {{ r}}^{4}
\\
&+ \left( {-\frac {7685}{
1179648}}{{ s}}^{2}+ \left( {\frac {1015}{3072}}{ y}+{
\frac {137}{16384}}{ x} \right) { s}-{\frac {6607}{2048}}
{{ y}}^{2}-{\frac {1071}{4096}}{ x}{ y}-{\frac{45}
{1024}}{ x}{ z}+{\frac {26757}{131072}}{{ x}}^{2}
 \right) {{ r}}^{3}
\\
 &+ \left[ {\frac {96835}{1179648}}{{ s}}^{
3}+ \left( {-\frac {1109}{65536}}{ x}+{\frac {281}{6144}}{
 y} \right) {{ s}}^{2}
+\left( {\frac {3}{512}}{ x}{ z}-{\frac {4593}{2048}}
{{ y}}^{2}+{\frac {20421}{131072}}{{ x}}^{2}-{\frac {915}{
4096}}{ x}{ y} \right) { s}
\right.
\\&
\quad\quad\left.
-{\frac {27}{128}}{ y}{ x}
{ z}+{\frac {4005}{256}}{{ y}}^{3}+{\frac {16083}{4096}}{
 x}{{ y}}^{2}-{\frac {99}{512}}{{ x}}^{3}-{\frac {
4671}{2048}}{ y}{{ x}}^{2}
+{\frac {1215}{4096}}{{
x}}^{2}{ z}-{\frac {135}{2048}}{ x}{{ z}}^{2}
\right] {{ r}}^{2}
\\
&+
 \left[ {\frac {23893}{73728}}{{ s}}^{4} +\left( {\frac {173}{1536}}{ y}-{\frac {34795}{196608}}
{ x} \right) {{ s}}^{3}+ \left( -{\frac {1287}{131072}}{{
 x}}^{2}-{\frac {57}{512}}{{ y}}^{2}+{\frac {5}{64}}{
 x}{ z}-{\frac {207}{4096}}{ x}{ y} \right) {{
 s}}^{2}
\right.
\\&
\quad\quad\left.
 + \left( -{\frac{
1431}{1024}}{ y}{{ x}}^{2}+{\frac {1107}{128}}{{ y}}
^{3}-{\frac {9}{64}}{ y}{ x}{ z}-{\frac {1143}{8192}
}{{ x}}^{3}+{\frac {4851}{2048}}{ x}{{ y}}^{2}-{
\frac {45}{1024}}{ x}{{ z}}^{2}+{\frac {405}{2048}}{{
 x}}^{2}{ z} \right) { s}
\right.
\\&
\quad\quad\left.
 -{\frac {81}{32}}{ y}{{
 x}}^{2}{ z}+{\frac {27}{16}}{ x}{ z}{{ y}}^
{2}
-{\frac {27}{128}}
{{ x}}^{3}{ z}+{\frac {2997}{1024}}{ y}{{ x}}^{3
}-{\frac {729}{16}}{{ y}}^{4}-{\frac {81}{4}}{ x}{{
 y}}^{3}-{\frac {6561}{16384}}{{ x}}^{4}+{\frac {19521}{
2048}}{{ x}}^{2}{{ y}}^{2} \right] { r}
\\
&+{\frac {88247}{
294912}}{{ s}}^{5}+ \left( {\frac {3113}{16384}}{ y}-{
\frac {133757}{786432}}{ x} \right) {{ s}}^{4}
+ \left( {\frac {347}{1536}}{{ y}}^{2
}+{\frac {1}{16}}{ x}{ z}-{\frac {18779}{1179648}}{{
 x}}^{2}+{\frac {141}{4096}}{ x}{ y} \right) {{ s}}
^{3}
\\&+
 \left( -{\frac {1147}{49152}}{
{ x}}^{3}-{\frac {9}{128}}{ y}{ x}{ z}+{\frac {
237}{256}}{{ y}}^{3}-{\frac {121}{2048}}{ y}{{ x}}^{2
}-{\frac {45}{2048}}{ x}{{ z}}^{2}+{\frac {5429}{12288}}
{ x}{{ y}}^{2}-{\frac {299}{4096}}{{ x}}^{2}{ z}
 \right) {{ s}}^{2}
 \\
 &+ \left( {\frac {7923}{2048}}{{ x}}^{2}{{
 y}}^{2}+{\frac {23}{32}}{{ x}}^{2}{{ z}}^{2}+{\frac {
1551}{1024}}{ y}{{ x}}^{3}-{\frac {3411}{16384}}{{ x
}}^{4}-{\frac {43}{32}}{ y}{{ x}}^{2}{ z}+{\frac {11}
{16}}{ x}{ z}{{ y}}^{2}-{\frac {47}{128}}{{ x
}}^{3}{ z}-{\frac {15}{2}}{ x}{{ y}}^{3}-{\frac {243}
{16}}{{ y}}^{4} \right) { s}
\end{split}
\end{gather}
}
Given the ``$v$-valued'' couplants $\{r, s, x, z\} = \{0.0109, 0.00324, 0.0253, 0.0329\}$, we obtain a
1\% increase in the predicted
Higgs mass (to $218\,{\rm GeV}$) within radiative electroweak symmetry breaking, as well as an
1\% increase in the extracted value of
the scalar field self-interaction couplant $y$ (to $y=0.0545$).
We further note that the method of characteristics approach leading to
Eq.\ (\ref{V_eff_char}) [and ultimately, to the series coefficients within Eq.\ (\ref{eq7.1})]
is gauge parameter independent to the
order we consider, since the specific one-loop RG functions entering the effective potential (\ref{V_eff_char})
are all gauge-parameter
independent. However, the gauge independence of results from the effective potential \cite{gauge1} in general rests
upon the implementation of Nielsen identities \cite{gauge2,gauge1}, whose applicability to radiatively
broken scenarios is problematical \cite{gauge2,gauge3}.

\subsection{\bf {\em The $z=0$ Case}}
In principle, the coupled equations (\ref{char_z})--(\ref{char_y}) could be solved numerically to provide $V_{eff}$ in
(\ref{V_eff_char}).  However, in the radiatively-broken theory, the initial condition for $y$ needed for a numerical solution must be
self-consistently  determined from the shape of the effective potential itself, a formidable numerical problem. Since the summation of
leading  logarithms contains the analytic dependence on $y$, such a summation facilitates the extraction of the self-consistent
solution for $y$ and the associated prediction of the Higgs mass.
These issues for a radiatively-broken theory should be contrasted with standard spontaneous symmetry breaking, where the relation
$y=m_\phi^2/(8\pi^2 v^2)$
 between $y$ and the Higgs mass can be exploited for a numerical exploration of the effective potential in the Higgs-mass parameter
 space \cite{quiros}.

 However, a closed form (method-of-characteristics) non-numerical solution to Eqs.\ (\ref{char_x})--(\ref{char_y}) does exist in the
z=0 limit.  This closed form solution is of particular value in determining the scale for new physics.
As already noted in Section 8.1, the singularity at $w=0$ characterizing each term of the series (\ref{eq6.1}), as evident in Eq.\
(\ref{eq8.5}), need not represent a true singularity of the leading logarithm sum.  To
confirm the need for an ${\cal O}(5\,{\rm TeV})$ scale for new physics within a radiative approach to the breakdown of electroweak
symmetry, we utilize the explicit solutions to Eqs.\ (\ref{char_x})--(\ref{char_y}) when the QCD couplant
$z$ is equal to 0:
\begin{gather}
\bar x(t)=\frac{x}{1-\frac{9}{4}xt},
\label{x_sol}
\\
\bar\phi(t)=\phi\left[1-\frac{9}{4}xt\right]^{1/3},
\label{phi_sol}
\\
\bar y(t)=\bar x(t)
\left[\frac{u_+\left(\frac{y}{x}-u_-\right)-u_-\left(\frac{y}{x}-u_+\right)\left(1-\frac{9}{4}xt\right)^{-\sqrt{65}/3}}
{\frac{y}{x}-u_--\left( \frac{y}{x} - u _ + \right) \left(1-\frac{9}{4}xt\right)^{-\sqrt{65}/3}}
\right] ,
\label{y_sol}
\end{gather}
where
\begin{equation}
u_\pm=\frac{-1\pm\sqrt{65}}{16}.
\label{u_pm}
\end{equation}
The $z=0$ effective potential, including the arbitrary $K\phi^4$ counterterm, is found from Eqs.\ (\ref{x_sol})--(\ref{y_sol}) to be
\begin{equation}
V=\pi^2\phi^4\left\{
K+xR^{1/3}
\left[\frac{u_+\left(\frac{y}{x}-u_-\right)R^{\sqrt{65}/3}-u_-\left(\frac{y}{x}-u_+\right)}
{\left(\frac{y}{x}-u_-\right)R^{\sqrt{65}/3}-\left(\frac{y}{x}-u_+\right)}
\right]\right\},~R\equiv1-\frac{9}{8}xL .
\label{z_zer_veff}
\end{equation}
Given that $x=g_t^2/4\pi^2=0.0253$ as before, we follow the procedure of the previous section by imposing the renormalization
conditions
\begin{equation}
V'(v)=0,~V^{(4)}(v)=24\pi^2y,
\end{equation}
at the electroweak vacuum expectation value $v=2^{-1/4}G_F^{-1/2}=246\,{\rm GeV}$, and find that $y=0.05403$ and $K=-0.05817$.  Upon
incorporation of these numbers into Eq.\ (\ref{z_zer_veff}), we obtain a potential that grows from its minimum at $\phi=v$ ({\it i.e.}
$R=1$) to $+\infty$ as $\phi$ approaches $16.5v\approx 4\,{\rm TeV}$ from below, {\it i.e.} as
\begin{equation}
R\to\left[
\frac{\left(\frac{y}{x}-u_+\right)}{\left(\frac{y}{x}-u_-\right)}
\right]^{3/\sqrt{65}}=0.8404
\label{R_limit}
\end{equation}
from above.  Thus, the $z=0$ exact solution to the one-loop RG equation is seen to be a potential that is bounded from below over the
domain for which it is valid.  The singularity which terminates this domain is assumed, as before, to set the scale for new physics at
a value not too different from the $5.5\,{\rm TeV}$ singularity seen previously to characterize each term in the series expansion
(\ref{eq6.1}).

\subsection{\bf {\em Zero-Field Limit of the Effective Potential}}
The zero-field limit of the effective potential $V_{LL} = \pi^2 \phi^4
(S_{LL} + K)$ corresponds to the large logarithm limit of the series
(\ref{eq6.1}) in which $L \rightarrow - \infty$ and $[(w-1)/w]^k \rightarrow 1$.
In this limit, the asymptotic behaviour of the leading contributions to
the series (\ref{eq6.1}) is seen from Eqs.\ (\ref{eq6.10}), (\ref{eq6.11}), (\ref{eq6.14}), (\ref{eq6.18}) and
(\ref{eq6.23})--(\ref{eq6.26}) to be
\begin{gather}
y F_0 = \frac{y}{w}
\begin{array}{c}{}\\_{\longrightarrow} \\ ^{|L| \rightarrow
\infty}
\end{array}
 - \frac{1}{3L},
\label{eq8.13}
\\
x F_1 = \frac{x M (zL)}{4} \left( \frac{w-1}{w} \right)^2
\begin{array}{c}{}\\_{\longrightarrow} \\ ^{|L| \rightarrow
\infty}
\end{array}
\frac{2}{L} \left( \frac{x}{z} \right),
\label{eq8.14}
\\
x^2 L F_2 = x^2 L \sum_{k=0}^3 H_k (z L) [(w-1)/w]^k
\begin{array}{c}{}\\_{\longrightarrow} \\ ^{|L| \rightarrow
\infty}
\end{array}
- \frac{3}{2L} \left( \frac{x}{z}\right)^2.
\label{eq8.15}
\end{gather}
For $n \geq 3$, the asymptotic behaviour of $x^n L^{n-1} F_n (w, \zeta)$
can be extracted from Eq.\ (\ref{eq6.7}) by noting from Eq.\ (\ref{eq8.5}) that dependence
of $F_n$ on the variable $w = 1 - 3y L$ disappears in the large
logarithm limit
\begin{equation}
F_n
\begin{array}{c}{}\\_{\longrightarrow} \\ ^{|L| \rightarrow
\infty}
\end{array}
\sum_{k=0}^{n+1} f_{n,k} (\zeta).
\label{eq8.16}
\end{equation}
In the large $w$ limit, we see from Eqs.\ (\ref{eq8.6b}) and (\ref{eq8.6c}) that we
can ignore all derivatives with respect to $w$ appearing on the
right-hand side of Eq.\ (\ref{eq6.7}).  Moreover, in the large $L$ limit,
the operators on the left hand side of Eq.\ (\ref{eq6.7}) that are
degree-1 in the variable $\zeta = zL$ dominate the remaining operators
[$w(w-1) d/dw$ is degree-zero via Eq.\ (\ref{eq8.6a})],
leading to the following recursion relation for $F_n$ in the large-$|L|$
limit:
\begin{equation}
\left( \frac{7}{4} \zeta^2 \frac{d}{d\zeta} + 2n \zeta \right) F_n =
\frac{3(3n-7)}{8} F_{n-1}, \; \; ( n \geq 3 ).
\label{eq8.17}
\end{equation}
We see from Eq.\ (\ref{eq8.17}) that $F_n$ is asymptotically one degree lower in
$\zeta$ than $F_{n-1}$.  Since the large-$|L|$ limit of $F_2$ is just $-
\frac{3}{2} \zeta^{-2}$ [Eq.\ (\ref{eq8.15})], we see from Eq.\ (\ref{eq8.17}) that $F_n$
in the large-$|L|$ limit is necessarily proportional to $\zeta^{-n}$,
\begin{equation}
\lim_{|L| \rightarrow \infty} F_n (w, \zeta) \equiv F_n (\zeta) = f_n
\zeta^{-n},
\label{eq8.18}
\end{equation}
such that
\begin{equation}
f_n = \frac{3(3n-7)}{2n} f_{n-1}, \; \; \; ( n \geq 3 ),
\label{eq8.19}
\end{equation}
with $f_2 = f_3 = - \frac{3}{2}$.  Since $\zeta = zL$, it is evident from
Eq.\ (\ref{eq8.18}) that each term in the series $S_{LL}$ (\ref{eq6.1}) goes like $1/L$ in
the large-$|L|$ limit,
\begin{equation}
x^n L^{n-1} F_n (w, \zeta)
\begin{array}{c}{}\\_{\longrightarrow} \\ ^{|L| \rightarrow
\infty}
\end{array}
\frac{f_n}{L} \left( \frac{x}{z} \right)^n,
\label{eq8.20}
\end{equation}
a structural result also upheld for $n = \{0,1,2\}$ in Eqs.\ (\ref{eq8.13}) -- (\ref{eq8.15}).
Thus {\it truncations of the series $S_{LL}$ vanish term-by-term as} $|L|
\rightarrow \infty$ ({\it i.e.} as $\phi \rightarrow 0$ or $\infty$), a
result that could hardly be anticipated from the form of $S_{LL}$
originally presented in Eq.\ (\ref{eq5.9}).

Indeed, it is clear from Eq.\ (\ref{eq8.20}) that the large-$|L|$ limit infinite
series (\ref{eq6.1}) for $S_{LL}$ can be summed explicitly if $|x/z|$ is
sufficiently small.  One easily sees from Eq.\ (\ref{eq8.19}) that $|x/z| < 2/9$
for convergence of $S_{LL}$ in this limit.  One can then make use of
Eqs.\ (\ref{eq8.13}) -- (\ref{eq8.15}) and (\ref{eq8.20}) within Eq.\ (\ref{eq6.1}) to obtain explicitly the
large-$|L|$ limit of the effective potential itself, which is governed
by the sum of its leading logarithm contributions:
\begin{equation}
V_{eff}
\begin{array}{c}{}\\_{\longrightarrow} \\ ^{|L| \rightarrow
\infty}
\end{array}
\frac{\pi^2 \phi^4}{L} \left[ - \frac{1}{3} + 2 \left(
\frac{x}{z}\right) - \frac{3}{2} \left( \frac{x}{z} \right)^2 + T \left(
\frac{x}{z} \right)\right],
\label{eq8.21}
\end{equation}
where
\begin{equation}
T(\rho) \equiv \sum_{n=3}^\infty \rho^n f_n.
\label{eq8.22}
\end{equation}
The summation $T(\rho)$, as defined by Eq.\ (\ref{eq8.22}), may be evaluated
from the recursion relation (\ref{eq8.19}), first by multiplying both sides of
Eq.\ (\ref{eq8.19}) by $2n \rho^{n-1}$, and then by summing from $n = 4$ to
$\infty$:
\begin{equation}
2 \sum_{n=4}^\infty n \rho^{n-1} f_n = 9\rho \sum_{n=4}^\infty (n-1)
\rho^{n-2} f_{n-1} - 12 \sum_{n=4}^\infty \rho^{n-1} f_{n-1}.
\label{eq8.23}
\end{equation}
Eq.\ (\ref{eq8.23}) is just a first order differential equation for $T(\rho)$, as
defined by Eq.\ (\ref{eq8.22}):
\begin{equation}
2 \left[ \frac{dT}{d\rho} - 3\rho^2 f_3 \right] = 9\rho \frac{dT}{d\rho}
- 12T
\label{eq8.24}
\end{equation}
with $f_3 = -3/2$ and with $T(0) = 0$.  When the solution to this
equation for $0 < \rho < 2/9$,
\begin{equation}
T(\rho) = \frac{1}{3} - 2\rho + \frac{3}{2} \rho^2 - \frac{1}{3} \left[
1 - \frac{9\rho}{2} \right]^{4/3},
\label{eq8.25}
\end{equation}
is substituted into Eq.\ (\ref{eq8.21}), one finds that
\begin{equation}
V_{eff}
\begin{array}{c}{}\\_{\longrightarrow} \\ ^{|L| \rightarrow
\infty}
\end{array}
- \frac{\pi^2 \phi^4}{3L} \left[ 1 - \frac{9x}{2z} \right]^{4/3}, \; \;
\; 0 \leq x/z \leq 2/9.
\label{eq8.26}
\end{equation}
Since $L = \log(\phi^2 / v^2) \longrightarrow -\infty$ as $\phi
\longrightarrow 0^+$, one immediately sees that $V_{eff} \longrightarrow
0^+$ as $\phi \longrightarrow 0^+$.  Thus if leading logarithms dominate the zero-field limit, then $\phi=0$ corresponds to {\it a
local minimum} of the effective potential when $0 \leq x/z \leq 2/9$, a
most surprising result.  In Appendix D, it is shown that Eq.\ (\ref{eq8.26})
cannot be extended past the $x/z < 2/9$ radius of convergence for the
series $T (x/z)$.  Thus, we come away knowing only that $\phi = 0$ is a
local minimum of the effective potential when QCD is sufficiently strong
($\alpha_s > 0.36$).  On the basis of (\ref{R_limit}), we cannot say anything about $\phi = 0$ being (or
not being) a local minimum if $\alpha_s$ is below this bound, as is the
case for the standard model parameters we have been using [$z = 0.0329 =
\alpha_s (v)/\pi$].\footnote{Strictly speaking the $z=0$ potential \protect(\ref{z_zer_veff}) also exhibits a local minimum at
$\phi=0$ [$R=+\infty$] immediately followed by a numerically tiny local maximum at a value $\phi=v\exp{(-2466)}$ [corresponding to
$R=141.4$] that is only infinitesimally separated from the $\phi=0$ local minimum.  Subject to the precision limitations of an actual
plot, which are insensitive to such small separations in the values of $\phi$, Eq.\ (\protect\ref{z_zer_veff}) is seen to exhibit an
apparent maximum at $\phi=0$, followed by its minimum at $\phi=v$ and its subsequent positive approach
to the singularity at $\phi =16.5 v$.  }

Nevertheless, this result may prove indicative of electroweak symmetry
restoration within a strong-phase context for QCD.  Recent work based
upon the ordering of Pad\'e-approximant zeros and poles for the
$\overline{MS}$ QCD $\beta$-function series \cite{Q} suggests that when
$n_f < 6$, the QCD couplant may exhibit the same two-phase behaviour
known to characterize the exact $\beta$-function for $N=1$
supersymmetric Yang-Mills theory (SYM), in which coexisting
strong-couplant and (asymptotically-free) weak-couplant phases evolve toward a
common infrared attractive point \cite{R}.  Within such a context, one may
envision a scenario in which the $\phi = 0$ minimum of preserved
electroweak symmetry is upheld by an effective potential involving the
strong phase of QCD, but in which a $\phi = v$ minimum of
radiatively-broken electroweak symmetry characterizes this same effective potential
when QCD is in its asymptotically-free weak phase.  Since the weak-phase
minimum at $\phi = v$ is deeper than the strong-phase minimum at $\phi =
0$ [$V_{eff} (v) < 0$ for $z = 0.033$, and $V_{eff} (0) = 0$ for all $z$],
the weak-phase of QCD is seen to be energetically preferred.
Thus, radiative electroweak symmetry breaking may also serve to select
the asymptotically-free phase of QCD over any coexisting strong phase.

\section*{Appendix A - The One-Loop Effective Action in Scalar QED}

\renewcommand{\theequation}{A.\arabic{equation}}
\setcounter{equation}{0}

In this appendix, we will compute the one-loop effective action 
in arbitrary covariant gauge for massless 
scalar quantum electrodynamics up to second order in the derivative
of the scalar field.

The classical Lagrangian for scalar QED with masses is
\begin{equation}
{\cal{L}}_{c1}= \Delta {\cal{L}} - \frac{m^2}{2} \left(\phi_1^2 + \phi_2^2 \right)- \frac{1}{4} F_{\mu\nu}
F_{\mu\nu} - \frac{\kappa^2}{2} (A_\mu + (1/\kappa)\partial_\mu S)^2,\qquad
(D_\mu = \partial_\mu - ie A_\mu, \; \; \; F_{\mu\nu} = 
\partial_\mu A_\nu - \partial_\nu A_\mu)
\label{eqA.1}
\end{equation}
where $\Delta {\cal{L}}$ is given by Eq.\ (\ref{eq1.2}) in terms of the real
scalar fields $\phi_1$, and $\phi_2$, which have been assigned equal
masses in Eq.\ (\ref{eqA.1}).  $A_\mu$ is a $U(1)$ vector and $S$ is a 
Stueckelberg scalar \cite{S}. For calculational convenience, we work in Euclidean
space $\left[ (\partial_\mu \phi_k)^2 \rightarrow - (\partial_\mu \phi_k)^2\right]$.
If $\Phi \equiv (\phi_1 + i \phi_2)/\sqrt{2}$, we note that 
${\cal{L}}_{c1}$ is invariant under the gauge transformation
\begin{gather}
\Phi \rightarrow e^{i\Lambda} \Phi
\label{eqA.2a}
\\
A_\mu \rightarrow A_\mu + \frac{1}{e} \partial_\mu \Lambda
\label{eqA.2b}
\\
S \rightarrow S - \frac{\kappa}{e} \Lambda
\label{eqA.2c}
\end{gather}
We now split $\phi_1$ and $\phi_2$  into the sum of background fields
$f_1, f_2$ and quantum fields $h_1, h_2$:
\begin{equation}
\phi = f_1 + h_1, \; \; \phi_2 = f_2 + h_2.
\label{eqA.3}
\end{equation}

The gauge fixing Lagrangian
\begin{equation}
{\cal{L}}_{gf} = -\frac{1}{2\alpha} \left[ \partial \cdot A + ie \alpha (f_1+if_2)(h_1 - ih_2) + \alpha \kappa S \right] 
\left[\partial \cdot A - ie \alpha (f_1 - if_2)(h_1 + ih_2) + \alpha \kappa S \right]
\label{eqA.4}
\end{equation}
is now employed.  
Upon setting the mass parameters $m^2$ and $\kappa^2$ equal to zero
(thereby decoupling $S$), the
terms in ${\cal{L}}_{c1} + {\cal{L}}_{gf}$ that
are bilinear in the quantum fields $A_\mu$, $h_1$ and $h_2$ are
\begin{gather}
 {\cal{L}}^{(2)} = - \frac{1}{2} V^t \hat{M} V,
\label{eqA.5}
\\
\hat{M} \!=\!
 \left(
\begin{array}{ccc}
p^2 + \frac{\lambda}{2} f_1^2 + \frac{\lambda}{6} f_2^2 + e^2 \alpha (f_1^2 + f_2^2)  & \lambda f_1 f_2 / 3 & 0 \\
\lambda f_1 f_2 / 3 & \!\!\!p^2 + \frac{\lambda}{6} f_1^2 + \frac{\lambda}{2} f_2^2 + e^2 \alpha (f_1^2 + f_2^2) & 0 \\
0 & 0 & \!\!\!p^2 (T_{\mu\nu} + \frac{1}{\alpha} L_{\mu\nu}) + e^2 (f_1^2 + f_2^2) \delta_{\mu\nu}
\end{array}
\right) 
\nonumber
\end{gather}
where $V = (h_1, h_2, A_\mu), \; \; T_{\mu\nu} = \delta_{\mu\nu} - p_\mu p_\nu / p^2 \;, \; L_{\mu\nu} = p_\mu p_\nu / p^2 \; \; \; \; 
(p=-i\partial)$.

We now expand the background fields $f_k$ so that
\begin{equation}
f_k (x) = \phi_k + x_\lambda \partial_\lambda \phi_k + \frac{1}{2}
x_\lambda x_\sigma \partial_\lambda \partial _\sigma \phi_k
\label{eqA.6}
\end{equation}
with higher order derivatives being neglected.

The one loop effective action is given by \cite{T} 
\begin{equation}
\Gamma^{(1)} = \ln \; {\det}^{-1/2} \hat{M}.
\label{eqA.7}
\end{equation}
This matrix $\hat{M}$ is then split into two parts,
$\hat{M} = H_0 + H_1$ where $H_0$ consists of these contributions to $\hat{M}$ that are independent of $\partial_\lambda \phi_k$
and $\partial_\lambda \partial_\sigma \phi_k$.  One now regulates $\Gamma^{(1)}$ in Eq.\ (\ref{eqA.7}) using the $\zeta$-function
\begin{equation}
\left. \Gamma^{(1)} = \frac{1}{2} \frac{d}{ds} \right|_{s=0} \left[
(\mu^2)^s \zeta(s)\right], 
\label{eqA.8}
\end{equation}	
where
\begin{equation}
\zeta(s) = \frac{1}{\Gamma(s)} \int_0^\infty dt t^{s-1} tr e^{-\hat{M}t}.
\label{eqA.9}
\end{equation}
Now $tr e^{-\hat{M}t}$ is expanded using the Schwinger formula \cite{V},
\begin{equation}
tr e^{-(H_0 + H_1)t}  =  tr \left\{ e^{-H_0 t} + \frac{(-t)}{1} e^{-
H_0 t} H_1 
 + \frac{(-t)^2}{2} \int_0^1 du \; e^{-(1-u) H_0 t} H_1 e^{-u H_0 t} H_1
+ \ldots \right\}.
\label{eqA.10}
\end{equation}
Upon diagonalizing the matrix (\ref{eqA.5}), we find the first contribution to
Eq.\ (\ref{eqA.10}) to be
\begin{equation}
tr e^{-H_0 t} = tr \exp - t \left( 
\begin{array}{ccc}
	p^2 + g_a \phi^2 & 0 & 0 \\
	0 & p^2 + g_b \phi^2 & 0 \\
	0 & 0 & p^2 (T_{\mu\nu} + \frac{1}{\alpha} L_{\mu\nu}) + g_A
\phi^2 \delta_{\mu\nu}
\end{array}
\right)
\label{eqA.11}
\end{equation}
where $g_a = \lambda/2 + e^2 \alpha, \; \; g_b = \lambda/6 + e^2 \alpha,
\; \; g_A = e^2$, with $\phi^2 \equiv \phi_1^2 + \phi_2^2$. 
Since $T_{\mu\nu}$ and $L_{\mu\nu}$ constitute a
complete set of orthogonal projection operators, Eq.\ (\ref{eqA.11}) reduces to
\begin{equation}
tr e^{-H_0 t}  =  tr \left\{ e^{-(p^2 + g_a \phi^2)t} + e^{-(p^2 + g_b
\phi^2)t} 
 +   \left(T_{\mu\nu} e^{-p^2 t} + L_{\mu\nu} e^{-p^2 t/\alpha} \right)
e^{-g_A \phi^2 t} \right\}.
\label{eqA.12}
\end{equation}
The term linear in $H_1$ in Eq.\ (\ref{eqA.10}) will not contribute to the one-loop
effective action
up to and including terms containing two derivatives of the background field.  From
the term in Eq.\ (\ref{eqA.10}) that is quadratic in $H_1$ we find that
\begin{equation}
\begin{split}
 &tr \; e^{-(1-u)H_0 t} H_1 e^{-u H_0 t} H_1
\\
\qquad =&  tr \Biggl\{e^{-(1-u)(p^2 + g_a \phi^2)t} (2 g_a \phi x \cdot \partial \phi) e^{-u(p^2 + g_a \phi^2) t} (2g_a \phi x \cdot \partial \phi) \Biggr. \\
& +  e^{-(1-u)(p^2+g_b \phi^2)r} (2g_b \phi x \cdot \partial \phi)
e^{-u(p^2+g_b \phi^2)t} (2g_b \phi x \cdot \partial \phi)
\\
& + \Biggl. \left( T_{\mu\nu} e^{-(1-u)p^2 t} + L_{\mu\nu} e ^{-(1-u)p^2
  t/\alpha} \right) (2g_A \phi x \cdot \partial \phi)
 \left( e^{-up^2 t} T_{\mu\nu} + e^{-up^2 t/\alpha}
L_{\mu\nu}\right) (2g_A \phi x \cdot \partial \phi) \Biggr\},
\end{split}
\label{eqA.13}
\end{equation} 
where $\phi x \cdot \partial \phi \equiv \sum_{k=1}^2 \phi_k x_\lambda
(\partial_\lambda \phi_k)$.  If the traces in Eq.\ (\ref{eqA.12}) and (\ref{eqA.13}) are computed
in momentum space, we see upon using the identities
\begin{equation}
tr \; e^{-p^2 t} = \int \frac{d^4 p}{(2\pi)^4} \; e^{-p^2 t} = \frac{1}{(4\pi t)^2}
\label{eqA.14}
\end{equation}
and
\begin{equation}
tr \left( f(p) x_\mu \; g(p) x_\nu \right) = \int \frac{d^4 p}{(2\pi)^4} \left( i \frac{\partial}{\partial p_\mu} f(p)\right)
\left( i \frac{\partial}{\partial p_\nu} g(p) \right)
\label{eqA.15}
\end{equation}
that
\begin{equation}
tr \; e^{-H_0 t} = \frac{1}{(4\pi t)^2} \left( e^{-g_a \phi^2 t} + e^{-g_b \phi^2 t} + (3+\alpha^2) e^{-g_A \phi^2 t} \right)
\label{eqA.16}
\end{equation}
and that 
\begin{equation}
\begin{split}
& \int_0^\infty dt \; t^{s-1} tr \left[ \frac{(-t)^2}{2} \int_0^1 du
    \; e^{-(1-u)H_0 t} H_1 e^{-u H_0 t} H_1\right]
\\
& =  2\phi^2 (\partial_\mu \phi)^2 \int_0^1 du \int
\frac{d^4 p}{(2\pi)^4} 
\\
&\qquad\Biggl\{ -u(1-u)p^2 \Gamma(s+4) \left(
\frac{g_a^2}{(p^2+g_a \phi^2)^{s+4}} 
 +  \frac{g_b^2}{(p^2+g_b \phi^2)^{s+4}} + \frac{3g_A^2}{(p^2+g_A
    \phi^2)^{s+4}} + \frac{g_A^2}{\left( \frac{p^2}{\alpha} + g_A
    \phi^2 \right)^{s+4}} \right)\Biggr.
\\
&\qquad \Biggl.+  \frac{3}{2} g_A^2 \frac{1}{p^2} \Gamma(s+2) \left( \frac{1}{(p^2 +
g_A \phi^2)^{s+2}} + \frac{1}{\left( \frac{p^2}{\alpha} + g_A \phi^2
\right)^{s+2}} 
 -   \frac{2}{\left[ \left( 1 - u + \frac{u}{\alpha} \right) p^2 + g_A
\phi^2 \right]^{s+2}} \right) \Biggr\}.
\end{split}
\label{eqA.17}
\end{equation}
The standard integral
\begin{equation}
\int \frac{d^n k}{(2\pi)^n} \frac{(k^2)^a}{(k^2 + M^2)^b} =
\frac{1}{(4\pi)^{n/2}}(M^2)^{n/2 + a - b}
\frac{\Gamma\left(a + \frac{n}{2}\right)}{\Gamma \left( \frac{n}{2}\right)}
\frac{\Gamma \left( b - a -\frac{n}{2} \right)}{\Gamma(b)}
\label{eqA.18}
\end{equation}
in conjunction with Eqs.\ (\ref{eqA.16}), (\ref{eqA.17}) and (\ref{eqA.9}) yields
\begin{equation}
\begin{split}
16\pi^2 \Gamma(s)\zeta(s)  =&   \left(g_a
\phi^2\right)^{2-s} + \left( g_b \phi^2 \right)^{2-s} + \left(
3+\alpha^2\right) \left( g_A \phi^2\right)^{2-s}
\\
& +  2\phi^2 \left( \partial_\mu \phi\right)^2 \int_0^1 du 
\Biggl[
-2 u (1-u) \Gamma (s+1) \Biggr.\\
&\qquad\qquad\qquad
\left( g_a^2 (g_a \phi^2)^{-1-s}
 +   g_b^2 \left( g_b \phi^2\right)^{-1-s} + 3g_A^2 \left( g_A
\phi^2\right)^{-1-s} + \alpha g_A^2 \left( g_A \phi^2\right)^{-1-s}
\right)\\
& \qquad\qquad\qquad\qquad\Biggl.+  \frac{3}{2} g_A^2 \Gamma (s+1) \left( \left(g_A
\phi^2\right)^{-1-s} 
+ \alpha \left( g_A \phi^2\right)^{-1-s} 
-   \frac{2}{1 - u + \frac{u}{\alpha}} \left( g_A
\phi^2 \right)^{-1-s} \right) \Biggr] ,
\end{split}
\label{eqA.19}
\end{equation}
leading via Eq.\ (\ref{eqA.8}) to the following one-loop effective action:
\begin{gather}
\begin{split}
32\pi^2\Gamma^{(1)}  = &  
\frac{1}{2} \left[ \left( g_a \phi^2 \right)^2 \left( \frac{3}{2} - \ln \frac{g_a \phi^2}{\mu^2} \right)
+ \left( g_b \phi^2\right)^2 \left( \frac{3}{2} - \ln \frac{g_b
  \phi^2}{\mu^2} \right)   
 +  (3+\alpha^2) \left( g_A \phi^2 \right)^2 \left( \frac{3}{2} -
 \ln \frac{g_A \phi^2}{\mu^2} \right) \right] 
\\
& +  2 \left( \partial_\mu \phi \right)^2 \left[ - \frac{1}{3} \left( g_a + g_b + (3+\alpha) g_A \right) 
 + \frac{3}{2} \left( g_A + \left( \alpha - \frac{2\alpha}{\alpha - 1} \ln \alpha \right) g_A \right) \right] 
\end{split}
\label{eqA.20}
\\
\left[ \phi^2 \equiv \phi_1^2 + \phi_2^2, \; \; (\partial_\mu \phi)^2 \equiv
(\partial_\mu \phi_1)^2 + (\partial_\mu \phi_2)^2\right].
\nonumber
\end{gather}
Further contributions to $\Gamma^{(1)}$ that contain more derivatives of the background 
field can also be computed.

The one-loop (1L) Landau gauge contribution to the effective potential
follows from the negative of the non-derivative terms in Eq.\ (\ref{eqA.20}) with
$\alpha$ taken to be equal to zero:
\begin{equation}
\begin{split}
\Delta V_{eff}^{1L}  = & \frac{1}{64\pi^2} \left( g_a^2 + g_b^2 + 3g_A^2 \right)
\left(\phi_1^2 + \phi_2^2 \right)^2 \left[
\log \frac{(\phi_1^2 +\phi_2^2)}{\mu^2} - \frac{3}{2}
 +  \frac{g_a^2 \log g_a + g_b^2 \log g_b + 3 g_A^2 \log
g_A}{g_a^2 + g_b^2 + 3g_A^2} \right]
\\
 = & \left( \frac{5\lambda^2}{1152 \pi^2} + \frac{3 e^4}{64 \pi^2}
\right) \left( \phi_1^2 + \phi_2^2 \right)^2 \left[ \log \frac{\phi_1^2 +
\phi_2^2}{\mu^2} + \mbox{constants} \right],
\end{split}
\label{eqA.21}
\end{equation}
a result consistent with Eq.\ (\ref{eq1.3}).  Since we have been using operator
regularization \cite{W}, no explicit renormalization is required to
excise divergences.  Nevertheless, finite counterterms may be added to
$\Gamma^{(1)}$ in Eq.\ (\ref{eqA.20}) to accommodate renormalization conditions
such as that of Eq.\ (\ref{eq1.4}), which is why the constants in Eq.\ (\ref{eqA.21})
remain unspecified;  e.g. under the renormalization condition (\ref{eq1.4}), the
usual numerical factor $-25/6$ would replace ``constants'' in the final
line of Eq.\ (\ref{eqA.21}).

The derivative terms in Eq.\ (\ref{eqA.20}) also contribute to the kinetic term
within the Euclidean-space effective action
\begin{equation}
\Gamma \equiv - V_{eff} - \frac{1}{2} Z_{(\phi)} \left[ (\partial_\mu
\phi_1)^2 + (\partial_\mu \phi_2)^2\right] + \ldots \; \; .
\label{eqA.22}
\end{equation}
Noting that $(\partial_\mu \phi_k)^2 \rightarrow - (\partial_\mu
\phi_k)^2$ in the Euclidean space version of Eq.\ (\ref{eq1.2}), we find from Eq.\
(\ref{eqA.20}) that
\begin{equation}
Z_{(\phi)} = 1 + \frac{g_a + g_b}{24\pi^2} - \frac{g_A}{8\pi^2} \left[
\frac{1}{2} + \alpha \left( \frac{7}{6} - \frac{3 \ln \alpha}{\alpha -
1}\right) \right],
\label{eqA.23}
\end{equation}
hence that
\begin{equation}
Z_{(\phi)} = 1 + \frac{\lambda}{36\pi^2} + \frac{e^2}{16\pi^2} +
{\cal{O}} (\lambda^2, e^4, \lambda e^2 )
\label{eqA.24}
\end{equation}
in Landau $(\alpha = 0)$ gauge.

\renewcommand{\theequation}{B.\arabic{equation}}
\setcounter{equation}{0}

\section*{Appendix B:  Scale-Invariant Scalar-Field Mass in ${\rm
M}\!\!\!$/SED}

For massless scalar electrodynamics, the requirement that $V$ be independent 
of the value of the scale parameter $\mu$ leads to the renormalization group 
equation \cite{A}
\begin{equation}
DV  \equiv \left[ \mu \frac{\partial}{\partial \mu} + \beta_\lambda (\lambda, e^2)
\frac{\partial}{\partial \lambda} + \beta_e (\lambda, e^2) \frac{\partial}{\partial
e^2}  
 +   \gamma(\lambda, e^2) \left( \phi \frac{\partial}{\partial \phi_1} + \phi_2 \frac{\partial}{\partial \phi_2} \right) \right]
V \left[ \mu, \lambda, e^2, \phi_k \right]=0
\label{eqB.1}
\end{equation}
with RG functions defined as in Eqs.\ (\ref{eq2.5}), (\ref{eq2.6}) and (\ref{eq2.7}). Taking derivatives of
Eq.\ (\ref{eqB.1}) with respect to scalar field components leads to
\begin{gather}
(D + \gamma) \frac{\partial V}{\partial \phi_k} = 0
\label{eqB.2}
\\
(D + 2\gamma) \frac{\partial^2 V}{\partial \phi_j \partial \phi_k} = 0
\label{eqB.3}
\end{gather}
These equations are upheld at all values of $\phi_k$, including the
vacuum expectation value (vev) defined by \\
$\partial V / \partial
\phi_k|_{\phi_{10}; \phi_{20}} = 0$.  Indeed this vev-defining equation
ensures that the vev components $\phi_{10}, \phi_{20}$ are themselves of
the functional form $\phi_{k0} = \phi_{k0} [\mu, \lambda(\mu), e^2 (\mu)]$.
Furthermore, since the anomalous field dimension associated with scalar
field components $\phi_k$ satisfies Eq.\ (\ref{eq2.5}), vev components
necessarily satisfy the equation
\begin{equation}
\left( \mu \frac{\partial}{\partial \mu} + \beta_\lambda
\frac{\partial}{\partial \lambda} + \beta_e \frac{\partial}{\partial
e^2} \right) \phi_{k0} = + \gamma \phi_{k0}.
\label{eqB.4}
\end{equation}
We also see from Eq.\ (\ref{eqA.22}) that
\begin{equation}
(D + 2\gamma) Z_{(\phi)} = 0.
\label{eqB.5}
\end{equation}

We now define the scalar-field mass as the square-root of the nonzero
eigenvalue of the matrix \\
$\hat{m}_\phi^2 \equiv \left[ \partial^2 V /
\partial \phi_j \partial \phi_k\right] / Z_{(\phi)}$ evaluated at vev
values of the scalar field component $(\phi_{10} \equiv \phi_0, \; \;
\phi_{20} = 0$).  This eigenvalue $(m_\phi^2)$ is a renormalization-group
invariant.  To see this, we consider
\begin{equation}
\begin{split}
\mu \frac{d m_\phi^2}{d\mu}  = & \left[ \mu \left( \frac{\partial \phi_0}{\partial \mu} \frac{\partial}{\partial \phi_0} + \frac{\partial}{\partial \mu}\right) + \beta_\lambda
\left( \frac{\partial \phi_0}{\partial \lambda} \frac{\partial}{\partial \phi_0} + \frac{\partial}{\partial \lambda} \right) + \beta_e
\left( \frac{\partial \phi_0}{\partial e^2} \frac{\partial}{\partial \phi_0} + \frac{\partial}{\partial e^2} \right) \right]\\
& \qquad\times\left[ \frac{V^{\prime\prime} [\mu, \lambda(\mu), e^2 (\mu), \phi_0
[\mu, \lambda(\mu), e^2 (\mu)]]}{Z_{(\phi)}[\mu, \lambda(\mu), e^2 (\mu), \phi_0 [\mu, \lambda(\mu), e^2 (\mu)]]}\right] .
\end{split}
\label{eqB.6}
\end{equation}
One easily sees from Eqs.\ (\ref{eqB.3}), (\ref{eqB.4}) and (\ref{eqB.5}) that $\mu d m_\phi^2 /
d\mu = 0$.  Note that $Z_{(\phi)}$ is only perturbatively removed from
unity [Eq.\ (\ref{eqA.23})], justifying the operational use of the second
derivative of the effective potential at the vev for calculating the
scalar field mass to a given order of perturbation theory.  We further
note that the RG-invariance of $V^{\prime\prime} / Z_{(\phi)}|_{\phi_0}$
has also been used to determine a nonperturbative $\beta$-function
within a toy $\phi_4^4$ model context \cite{X}.

\section*{Appendix C:  Solutions for $F_3 (w, \zeta)$}

\renewcommand{\theequation}{C.\arabic{equation}}
\setcounter{equation}{0}

$F_3 (w, \zeta)$ is given by Eq.\ (\ref{eq6.28}), with the individual $N_k
(\zeta)$ constituting solutions of Eqs.\ (\ref{eq6.29})--(\ref{eq6.33}) that are not
singular at $\zeta = 0$.  Thus Eq.\ (\ref{eq6.29}) is just the inhomogeneous
linear first order differential equation
\begin{equation}
\frac{d N_0}{d\zeta} + \frac{8(1+3\zeta)}{\zeta(7\zeta+4)} N_0 = \frac{3 H_0 
(\zeta)}{\zeta(7\zeta + 4)}
\label{eqC.1}
\end{equation}
with $H_0 (\zeta)$ given by Eq.\ (\ref{eq6.23}).  The integrating factor for Eq.\ (\ref{eqC.1}) is
\begin{equation}
g(\zeta) = \zeta^2 (\zeta + 4/7)^{10/7}.
\label{eqC.2}
\end{equation}
The only solution to Eq.\ (\ref{eqC.1}) that is not singular as $\zeta \rightarrow 0$ is
\begin{equation}
N_0 (\zeta)  =  \frac{3 \int_0^\zeta d t \; g(t) H_0 (t) / [t(7t+4)]}{g(\zeta)}
 =  \frac{\zeta^{-2}}{10} \left[ 10 \left( 1 + \frac{7\zeta}{4} \right)^{-9/7} - 
9\left( 1 + \frac{7\zeta}{4} \right)^{-10/7} - 1 \right].
\label{eqC.3}
\end{equation}
If one expands Eq.\ (\ref{eqC.3}) about $\zeta = 0$, one obtains Eq.\ (\ref{eq6.39}).

Corresponding solutions of Eqs.\ (\ref{eq6.30}) -- (\ref{eq6.33}) are listed below:
\begin{gather}
N_1 (\zeta)  =  \zeta^{-3} \left[ \frac{21\zeta}{5} - \frac{8}{5} + 12 \left( 1 
+ \frac{7\zeta}{4} \right)^{-2/7}
-  \frac{8}{3} \left( 1 + \frac{7\zeta}{4} \right)^{6/7} -
\frac{116}{15} \left( 1 + \frac{7\zeta}{4} \right)^{-3/7} \right],
\label{eqC.4}
\\
\begin{split}
N_2 (\zeta)  = & \zeta^{-4} \left[ - \frac{81}{10} \zeta^2 +
\frac{2023}{30} \zeta + \frac{1273}{30} 
+ \frac{64}{21} \left( 1 + \frac{7\zeta}{4} \right)^{13/7} -
\frac{831}{7} \left( 1 + \frac{7\zeta}{4} \right)^{6/7} \right.
\\
& +  \left. \frac{738}{7} \left( 1 + \frac{7\zeta}{4} \right)^{5/7} - \frac{16}{21}
\left( 1 + \frac{7\zeta}{4} \right)^{-2/7}
-  \frac{943}{30} \left( 1 + \frac{7\zeta}{4} \right)^{4/7} \right],
\end{split}
\label{eqC.5}
\\
\begin{split}
N_3 (\zeta)  = & \zeta^{-5} \left[ 4\zeta^3 - \frac{418}{3} \zeta^2 - \frac{524}{3} \zeta - \frac{160}{3} 
- \frac{128}{147} \left( 1 + \frac{7\zeta}{4} \right)^{20/7} +
\frac{7440}{49} \left( 1 + \frac{7\zeta}{4} \right)^{13/7}\right. 
\\
& +  \left. \frac{704}{49} \left( 1 + \frac{7\zeta}{4} \right)^{6/7} - \frac{1136}{7} \left( 1 + \frac{7\zeta}{4} \right)^{12/7}
- \frac{176}{21} \left( 1 + \frac{7\zeta}{4} \right)^{5/7} +
\frac{176}{3} \left( 1 + \frac{7\zeta}{4} \right)^{11/7}
\right],
\end{split}
\label{eqC.6}
\end{gather}
\begin{equation}
\begin{split}
N_4 (\zeta)  = & \zeta^{-6} \left[ 72 \zeta^3 + 144 \zeta^2 + 96 \zeta + \frac{64}{3} 
- \frac{6}{7} \left\{ \frac{384}{7} \left( 1 + \frac{7\zeta}{4}
\right)^{20/7} + \frac{128}{7} \left( 1 + \frac{7\zeta}{4}
\right)^{13/7}\right. \right.
\\
& +  \frac{32}{21} \left( 1 + \frac{7\zeta}{4} \right)^{6/7} - 64 \left( 1 + \frac{7\zeta}{4} \right)^{19/7}
-  \left. \left. \frac{32}{3} \left( 1 + \frac{7\zeta}{4} \right)^{12/7}
+ \frac{224}{9} \left( 1 + \frac{7\zeta}{4} \right)^{18/7} \right\}
\right].
\end{split}
\label{eqC.7}
\end{equation}

\section*{Appendix D:  Ambiguity of Zero-Field Limit when $x/z > 2/9$}

\renewcommand{\theequation}{D.\arabic{equation}}
\setcounter{equation}{0}

One must be careful not to extend the result (\ref{eq8.26}) past the radius of 
convergence of the series $T\left( \frac{x}{z}\right)$.  
Naively, the result (\ref{eq8.26}) would imply $V_{eff} \rightarrow 0^+$
{\it regardless} of $x/z$, since $[1 - 9x/2z]^{4/3}$ is positive-definite.
However, the solution (\ref{eq8.25}) for $T(\rho)$ is referenced to the initial
condition $T(0) = 0$.  If one assumes one can extend Eq.\ (\ref{eq8.22}) past its
$\rho = 2/9$ radius of convergence, one must again solve Eq.\ (\ref{eq8.24})
subject to the new initial condition $T(2/9) = -1/27$, 
as obtained from Eq.\ (\ref{eq8.25}) when $\rho = 2/9$.  
The most general solution to the differential equation
(\ref{eq8.24}) subject  to this new initial condition is 
\begin{equation}
T(\rho) = \frac{1}{3} - 2\rho + \frac{3}{2} \rho^2 + c (\rho -
2/9)^{4/3}, \; \; \; ( \rho > 2/9 ) , 
\label{eqD.1}
\end{equation}
where the constant $c$ is {\it arbitrary}.  If one now substitutes Eq.\
(\ref{eqD.1}) into Eq.\ (\ref{eq8.21}), the sign of $V_{eff}$ remains undetermined,
\begin{equation}
V_{eff} 
\begin{array}{c}{}\\_{\longrightarrow} \\ ^{|L| \rightarrow
\infty}
\end{array}
\frac{\pi^2 \phi^4 c}{L} \left[ \frac{x}{z} - \frac{2}{9} \right]^{4/3},
\; \; \; \frac{x}{z} > \frac{2}{9},
\label{eqD.2}
\end{equation}
reflecting the ambiguity in attempting to extend the 
determination of a series past its known radius of
convergence.

\section*{Acknowledgements}

We are grateful for discussions with L. Smolin and V. A. Miransky, as
well as for research support from the Natural Sciences and Engineering
Research Council of Canada.

\end{document}